\documentclass[11pt,a4paper]{article}
\pdfoutput=1
\usepackage{amsmath,amssymb,dsfont}
\usepackage{graphicx,color}
\usepackage{verbatim, cancel}
\usepackage[bf]{caption}
\usepackage{ulem,stackrel}

\usepackage{epsfig}

\usepackage[breaklinks,bookmarks=true,colorlinks=true,linkcolor=blue,citecolor=blue,urlcolor=blue,hyperfigures=true]{hyperref}
\usepackage[nosort]{cite} 
\usepackage[bulletsep]{collref} 

\numberwithin{equation}{section}

\ifx\href\asklfhas\newcommand{\href}[2]{#2}\fi

\definecolor{pink}{rgb}{0.7,0,0.7}
\definecolor{green}{rgb}{0,0.5,0}
\definecolor{orange}{rgb}{1,0.4,0.3}

\newcommand{\MH}[1]{{\color{pink}{#1}}}

\newcommand{\be}{\begin{equation}}
\newcommand{\ee}{\end{equation}}
\newcommand{\ba}{\begin{aligned}}
\newcommand{\ea}{\end{aligned}}
\newcommand{\ben}{\begin{displaymath}}
\newcommand{\een}{\end{displaymath}}
\newcommand{\bea}{\begin{eqnarray}}
\newcommand{\eea}{\end{eqnarray}}
\newcommand{\bean}{\begin{eqnarray*}}
\newcommand{\eean}{\end{eqnarray*}}
\newcommand{\bpmat}{\begin{pmatrix}}
\newcommand{\epmat}{\end{pmatrix}}

\newcommand{\AdS}{{\ensuremath{\text{AdS}}}}
\newcommand{\Sph}{{\ensuremath{\text{S}}}}
\newcommand{\EAdS}{{\ensuremath{\text{EAdS}}}}
\newcommand{\HS}{{\ensuremath{\text{H}}}}

\newcommand{\SO}{{\ensuremath{\text{SO}}}}

\newcommand{\alg}[1]{\ensuremath{\mathfrak{#1}}}
\newcommand{\as}{\ensuremath{\mathfrak{s}}}
\newcommand{\ah}{\ensuremath{\mathfrak{h}}}

\newcommand{\psu}{{\ensuremath{\mathfrak{psu}}}}
\newcommand{\so}{{\ensuremath{\mathfrak{so}}}}

\newcommand{\Integers}{{\ensuremath{\mathbb{Z}}}}
\newcommand{\Reals}{{\ensuremath{\mathbb{R}}}}

\newcommand{\calF}{{\ensuremath{\mathcal{F}}}}
\newcommand{\calG}{{\ensuremath{\mathcal{G}}}}

\newcommand{\ellF}{{\ensuremath{\text{F}}}}
\newcommand{\ellE}{{\ensuremath{\text{E}}}}
\newcommand{\ellPi}{{\ensuremath{\Pi}}}
\newcommand{\ellK}{{\ensuremath{\text{K}}}}

\renewcommand{\th}{\theta}

\renewcommand{\d}{\delta}
\newcommand{\eps}{\epsilon}

\newcommand{\s}{\sigma}

\newcommand{\vp}{\varphi}

\newcommand{\m}{\mu}

\newcommand{\vk}{\varkappa}

\newcommand{\Omt}{{\tilde{\Omega}}}

\newcommand{\pit}{{\tilde{\pi}}}
\newcommand{\phit}{{\tilde{\phi}}}

\newcommand{\rt}{{\tilde{r}}}
\newcommand{\zt}{{\tilde{z}}}

\newcommand{\p}{\partial}

\newcommand{\diag}{\ensuremath{\mathrm{diag}}}

\newcommand{\PP}{{\ensuremath{p}}}
\newcommand{\kk}{{\ensuremath{k}}}
\newcommand{\jj}{{\ensuremath{j}}}

\newcommand{\mm}{{\ensuremath{m}}}
\newcommand{\cc}{{\ensuremath{n}}}

\newcommand{\uu}{{\ensuremath{\sqrt{1+\vk^2}\,\alpha}}}

\newcommand{\dd}{{\ensuremath{\text{d}}}}
\newcommand{\abs}[1]{{|{#1}|}}
\newcommand{\ord}{{\ensuremath{\mathcal{O}}}}

\newcommand{\nn}{\nonumber}

\long\def\symbolfootnote[#1]#2{\begingroup
\def\thefootnote{\fnsymbol{footnote}}\footnote[#1]{#2}\endgroup}

\def\ads{{\rm AdS}_5\times {\rm S}^5}

\begin{document}

\begin{titlepage}
\hfill\parbox[t]{4cm}{\texttt{NORDITA-2016-93 \\
UUITP-16/16 \\ 
ZMP-HH/16-20}}
\vspace{20mm}

\begin{center}

{\Large \bf Superintegrability of Geodesic Motion on the 
Sausage Model}

\vspace{30pt}

\normalsize
{Gleb Arutyunov$^{a,}$\footnote{Correspondent fellow at Steklov Mathematical Institute, Moscow.}, Martin Heinze$^{a}$ and Daniel Medina-Rincon$^b$
}
\\[6mm]

{\small
{\it\ ${}^a$Institut f{\"u}r Theoretische Physik, 
	Universit{\"a}t Hamburg,\\ 
	Luruper Chaussee 149, 22761 Hamburg, Germany}\\[2mm]
{\it\ ${}^a$Zentrum f{\"u}r Mathematische Physik,
	Universit{\"a}t Hamburg,\\
	Bundesstrasse 55, 20146 Hamburg, Germany}\\[2mm]
{\it\ ${}^b$Nordita, KTH Royal Institute of Technology and Stockholm University,\\
	Roslagstullsbacken 23, SE-106 91 Stockholm, Sweden}\\[2mm]
{\it\ ${}^b$Department of Physics and Astronomy,
	Uppsala University,\\
	SE-751 08 Uppsala, Sweden}\\[5mm]
\texttt{gleb.arutyunov@desy.de},\qquad \texttt{martin.heinze@desy.de},\vphantom{\{}\\[.7mm]
\texttt{d.r.medinarincon@nordita.org}
}

\vspace{35pt}

\end{center}

\centerline{{\bf{Abstract}}}
\vspace*{5mm}
\noindent
\small
Reduction of the $\eta$-deformed sigma model on $\ads$ to the two-dimensional squashed sphere $({\rm S}^2)_{\eta}$ can be viewed as a special case of the Fateev 
sausage model where the coupling constant $\nu$ is imaginary. We show that geodesic motion in this model is described by a certain superintegrable mechanical system
with four-dimensional phase space.
This is done by means of explicitly constructing three integrals of motion which satisfy the $\alg{sl}(2)$ Poisson algebra relations, albeit being non-polynomial in momenta. Further, we find a canonical transformation 
which transforms the Hamiltonian of this mechanical system to the one describing the geodesic motion on the usual two-sphere. By inverting this transformation we map
geodesics on this auxiliary two-sphere back to the sausage model. {\it This paper is a tribute to the memory of Prof. Petr Kulish}.

\end{titlepage}

\newpage

{\hypersetup{linkcolor=black}
\tableofcontents}


\section{Introduction}
The appearance of integrable models in the context of the gauge-string correspondence  
 is a fascinating phenomenon\cite{Arutyunov:2009ga, Beisert:2010jr, Bombardelli:2016rwb}.  
In particular, the string sigma model on ${\rm AdS}_5\times {\rm S}^5$ in  the light-cone gauge defines a two-dimensional
integrable quantum field theory which spectrum is accessible through the mirror Thermodynamic Bethe Ansatz (TBA) \cite{Arutyunov:2009ur,Bombardelli:2009ns,Gromov:2009tv,Gromov:2009bc} and its alternative representation in terms of the Quantum Spectral Curve \cite{Gromov:2013pga}.

Recently, based on the previous work \cite{Klimcik:2002zj,Klimcik:2008eq}, it was understood that the string sigma model on $\ads$ can be deformed in a way which preserves its integrability. Notably, the deformations\footnote{The corresponding dual theories are expected to be described as non-commutative gauge theories, various aspects of the latter have been extensively studied in the work of P. Kulish and his collaborators, see {\it e.g.} \cite{Chaichian:2004za}.}  
come in three families which are typically called as $\eta$-deformations \cite{Delduc:2013qra,Delduc:2014kha}, $\lambda$-deformations \cite{Sfetsos:2013wia,Hollowood:2014qma} and deformations based on solutions of the classical Yang-Baxter equation \cite{Kawaguchi:2014qwa, Matsumoto:2014nra, Matsumoto:2014gwa, vanTongeren:2015soa, vanTongeren:2015uha}. This discovery triggered a number of interesting investigations which range from constructing the world-sheet S-matrices \cite{Arutyunov:2013ega, Hollowood:2015dpa, Hoare:2015kla} and the TBA equations  \cite{Arutyunov:2012zt,Arutyunov:2012ai,Arutynov:2014ota} for these theories to finding the 
corresponding background geometries \cite{Hoare:2014pna, Lunin:2014tsa, Hoare:2015gda, Arutyunov:2014cra, Arutyunov:2015qva, Hoare:2015wia, Arutyunov:2015mqj, Kyono:2016jqy, Hoare:2016hwh, Orlando:2016qqu, Borsato:2016zcf, Chervonyi:2016ajp, Borsato:2016ose}.

In the present work we will deal with yet another aspect of deformed theories, namely, with finite-dimensional integrable models which arise upon certain consistent reductions of the $\eta$-deformed models. 
In our previous work  \cite{Arutyunov:2014cda} and \cite{Arutyunov:2016ysi} we have shown that the ``spinning string ansatz" and its generalisation lead to a reduction of the 
original sigma model to one-dimensional models. It is natural  to refer to these models as  the $\eta$-deformed Neumann and Neumann-Rosochatius models, respectively, in accord with their undeformed counterparts appearing in string theory on 
$\ads$ \cite{Arutyunov:2003uj,Arutyunov:2003za}. For these mechanical models we proved their Liouville integrability by exhibiting a necessary number of commuting integrals of motion and by constructing the corresponding Lax representations. 
In particular, in  \cite{Arutyunov:2014cda} restricting ourselves for simplicity to the deformed five-sphere, we have demonstrated that the geodesic motion on this manifold is integrable -- in addition to the three conserved angular momenta and 
the Hamiltonian $H$, there is yet another non-trivial integral $Q$  which has vanishing Poisson brackets with the previous ones and which is quartic in momenta. The system which describes this motion 
is the $\eta$-deformed Rosochatius model.

In this paper we continue to investigate the geodesic problem on the $\eta$-deformed manifolds.\footnote{Special cases for geodesic motion in $\eta$-deformed backgrounds have been approached earlier in \cite{Kameyama:2014via} and in the unpublished work by S. Frolov and R. Roiban.} 
For the case of the usual sphere the geodesic motion is described by the Rosochatius model, which is known to be not only Liouville integrable but 
also maximally superintegrable \cite{Moser1980}. This means that it has a set of non-abelian integrals of motion, {\it viz.}, forming a non-trivial Poisson algebra. This is precisely the reason why geodesics on the sphere are closed. In particular, for the case of the five-sphere the Rosochatius model has a four-dimensional phase space 
(after accounting that three angle variables are cyclic and that their corresponding conserved angular momenta can be fixed) and there are three non-abelian integrals which make this model superintegrable. Of course, for a particle on a $d$-dimensional sphere the maximal superintegrability  follows from existence of the non-abelian isometry group ${\rm SO}(d+1)$.

Trials reveal that in general the question about superintegrability of geodesic motion for the $\eta$-deformed case is difficult to answer. In this work we show however that there exists a consistent reduction 
of the $\eta$-deformed sigma model for which the corresponding geodesic problem is superintegrable. Omitting the AdS part of the $(\ads)_{\eta}$ and putting to zero the three angular coordinates
$\xi$, $\phi_1$, and $\phi_2$ of $(\Sph^5)_\eta$ we obtain a reduced sigma model which depends on the remaining two coordinates, namely, $z$ and $\phi$
with the range $0\leq z \leq 1$ and $0\leq \phi<2\pi$. As has been noticed in \cite{Hoare:2014pna}, this reduced sigma model is nothing else but the Fateev sausage model \cite{Fateev:1992tk},
although for a real deformation parameter $\eta$ the corresponding manifold looks like a ``squashed sphere'' rather than a stretched out sausage, see Figure \ref{fig:FSplot}.
We will therefore attribute to this manifold the notation $(\Sph^2)_\eta$  but continue to call it Fateev sausage (FS). At $\eta=0$ the corresponding manifold is the usual two-sphere ${\rm S}^2$ which gets squashed as the deformation is turned on. A  very interesting feature is that when $\eta\to 1$ the model undergoes a topological transition so that for $\eta=1$ the corresponding manifold 
is a two-dimensional hyperbolic space ${\rm H}^2$ which coincides with the Euclidean version of the Anti-de-Sitter space. 

One of the main findings of our work is  that the geodesic problem in the FS model is maximally superintegrable. The point-like solutions describing geodesic motion have four-dimensional phase space.
The corresponding Hamiltonian coincides with that of the two-dimensional $\eta$-deformed Rosochatius model obtained from the five-dimensional one by
putting the angular coordinates $\xi, \phi_1,\phi_2$ and their corresponding momenta to zero\footnote{See the formula (A.1) in \cite{Arutyunov:2014cda}.  }.  This model appears to be maximally superintegrable because as we will show
 there are three integrals $K_A$, $A=1,2,3$, which satisfy the Poisson bracket relations 
for the generators of the $\alg{sl}(2)$ algebra with the Hamiltonian being proportional to the quadratic Casimir. 
Furthermore, $K_A$ also satisfy certain reality conditions which depend on the value of $\eta$. For $\eta=0$ the generators $K_A$
actually produce an $\so(3)$ algebra, while for $\eta=1$ the corresponding algebra is $\so(1,2)$, in agreement with the topological transition from ${\rm S}^2$ to ${\rm H}^2$
that was mentioned above. To find $K_A$ we first start from a perturbative treatment in powers of $\varkappa=\frac{2\eta}{1-\eta^2}$ as we know the form of the generators for $\varkappa=0$. But later 
we arrive at a general ansatz for $K_A$ which does not rely on perturbation theory anymore and reduces the problem of finding $K_A$ to a system of ODE's, which we solve explicitly.

The fact that for all values of $\varkappa$ the Poisson algebra of $K_A$ looks formally as $\alg{sl}(2)$ hints that there could be a non-trivial canonical transformation 
from the phase space $\{r,\phi , \pi_r,\pi_{\phi}\}$ to the one of the ordinary sphere $\{\tilde{r},\tilde{\phi},\tilde{\pi}_r,\tilde{\pi}_{\phi}\}$,  which transforms the Hamiltonian $H$
into the Hamiltonian describing the geodesic motion on the sphere. Finding of this canonical transformation is another achievement of our work.

Applying the inverse canonical transformation we obtain the geodesics in the FS model from the geodesics in the sphere.\footnote{Alternatively, geodesics in the FS model can be obtained directly by separation of variables. One just needs to be careful to pick up a real, {\it i.e.} physical, solution for the variable $r$, {\it cf.} formula (\ref{eq:IT-rinv}).} For this we observe 
another very interesting feature. Namely, while geodesics on ${\rm S}^2$ are closed due to superintegrability,
their images on $({\rm S}^2)_{\eta}$ are {\it not closed} for generic $\eta$, where we however determine the condition when they become closed.  At first sight this contradicts to superintegrability of  $({\rm S}^2)_{\eta}$. One might find an explanation of this phenomenon
by recalling the conditions under which periodic motion takes place. According to the corresponding theorem, see {\it e.g.} \cite{Perelomov:1990}, having $2n-1$ integrals for a dynamical system with 
$2n$ dimensional phase space will lead to a one-dimensional Liouville torus and therefore to closed orbits if the integrals are smooth functions of the phase 
space variables and their level surface is compact and connected. For generic $\eta$, this appears not be the case for the found $K_A$  -- the generators $K_1$ and $K_2$ contain inverse powers of momenta and therefore are not globally 
defined on the phase space. The points where $K_A$  have singularities must be excluded which, from this point of view, makes $({\rm S}^2)_{\eta}$ a much more complicated manifold. 
The situation is also somewhat similar to the one encountered  for Euler's top. Here the angular velocity $\vec{\Omega}$ in the body, {\it i.e.} in the rotating frame, undergoes a periodic motion, while 
the three Euler angles which describe the motion of the body in a stationary frame have in general  two rationally incomparable frequencies, the latter depend on the moments of inertia. 
It is this complicated map from $\vec{\Omega}$ to Euler angles which destroys periodicity \cite{Whittaker:1988}.

Finally, we point out that the developed methods for finding the integrals of motion and solving the geodesic problem 
are universal and apply to any two dimensional surface of revolution, having one isometry, which seems to connect to the classic works \cite{Legendre:1806,Oriani,Bessel,Bessel2}. 
As an example, for the renowned geodesics problem on the spheroid, {\it i.e.}, the ellipsoid of revolution, we provide the corresponding integrals of motion in an appendix.

The paper is organised as follows. The next section is devoted to some preliminaries and setting up notations. In Section \ref{sec:PIFS} we construct the integrals $K_A$ for generic values of 
$\eta$ and consider their values in the  limiting cases to $\Sph^2$ and $\HS^2$, $\eta\to 0$ and $\eta\to 1$. We also verify that they form the $\alg{sl}(2)$ algebra. In Section \ref{sec:NCC} we determine a canonical transformations
which maps $H$ to the standard Hamiltonian for geodesic motion on the two-sphere. In Section \ref{sec:IT} we find the corresponding inverse transformation and apply it to solve the geodesic problem for 
the FS model. Finally, in the outlook, Section \ref{sec:Outlook}, we briefly discuss the problems for future research. In Appendix \ref{app:Spheroid} we apply our methods to the case of the spheroid.

\section{Preliminaries and setup}
	The aim of this work is to study geodesic motion on the $\eta$-deformed two-sphere $(\Sph^2)_\eta$, that is the sphere part of the background obtained by $\eta$-deforming\cite{Delduc:2013qra,Delduc:2014kha,Arutyunov:2013ega} the $\AdS_2 \times \Sph^2$ superstring sigma model \cite{Berkovits:1999zq}. The construction has been carried out explicitly in \cite{Hoare:2014pna}, giving for the (bosonic) sphere part of the sigma model action and Lagrangian
	\be
		\label{eq:FS-sigmaLag}
		S_{(\Sph^2)_\eta} = T \!\int \dd^2\s L_{(\Sph^2)_\eta}~,\qquad
		L_{(\Sph^2)_\eta} = \frac{1}{2}\left(\frac{\p^\m z \p_\m z}{(1-z^2)(1+\vk^2 z^2)} + \frac{1-z^2}{1+\vk^2 z^2}\,\p^\m \phi \p_\m \phi\right)~,
	\ee
	with string tension $T$, deformation parameter $\vk=\frac{2\eta}{1-\eta^2}$, and $\eta$ the original parameter of \cite{Delduc:2013qra,Delduc:2014kha}.
	Furthermore, \eqref{eq:FS-sigmaLag} constitutes a consistent truncation of the $(\AdS_5 \times \Sph^5)_\eta$ Lagrangian \cite{Arutyunov:2013ega} by switching off the $(\AdS_5)_\eta$ degrees of freedom as well as the $\xi$-, $\phi_1$-, and $\phi_2$-directions on $(\Sph^5)_\eta$.
	
	As appreciated by the authors of \cite{Hoare:2014pna}, this is nothing but the Lagrangian of the well-known integrable Fateev sausage model \cite{Fateev:1992tk},
	\be
		L_{\rm FS} = \frac{1}{2}\left(
			\frac{ \p_\m \vec{n} \cdot \p^\m \vec{n}}{1 - \frac{\nu^2}{2\,g^2} n_3{}^2} + i\,g\,\th\,\mathcal{T} \right)
			\qquad\text{with}\quad \vec{n}^2 = \sum^3_{A=1}n_A{}^2 =1~,\quad 
			g=\frac{\nu}{2} \coth\left(\frac{\nu(t_0-t)}{4\pi}\right)~,
	\ee
	where the instanton charge $\mathcal{T}$ and the topological angle $\th$ do not concern us here. Especially, taking
	\be
		\vec{n} = \left(\sqrt{1-z^2} \cos(\phi),\sqrt{1-z^2} \sin(\phi),z\right)\,,
	\ee
	we recover \eqref{eq:FS-sigmaLag} by identifying the deformation parameters as
	\be
		\label{eq:FS-vknu}
		\vk = \pm\frac{i\,\nu}{\sqrt{2}\,g}~.
	\ee
	Although for $\vk \in \Reals$ the $(\Sph^2)_\eta$ background corresponds rather to an oblate squashed sphere than a prolate sausage, see Figure \ref{fig:FSplot} below, in agreement with \cite{Hoare:2014pna} we will refer to the corresponding $d=2$ dimensional manifold as the Fateev sausage (FS).

	Restriction to geodesic motion is obtained by considering a point-particle solution, $z=z(\tau)$ and $\phi=\phi(\tau)$ with $\tau$ being the world-sheet time, yielding for the Hamiltonian
	\be
		\label{eq:FS-hamz} 
		H = 
			\frac{(1-z^2)(1 + \vk^2 z^2)}{2}\pi_z^2
			+ \frac{1 + \vk^2 z^2}{2(1-z^2)}\pi_\phi^2~,
	\ee
	where $\pi_z$ and $\pi_\phi$ are the momenta canonically conjugate to $z$ and $\phi$, respectively. By cyclicity of $\phi$, we see that $\pi_\phi$ is an integral of motion,
	\be
		\{H,\pi_\phi\}_{\rm PB}=0~,
	\ee
	showing the Liouville integrability of the system.

	Let us additionally perform the canonical transformation given by
	\be
		\label{eq:FS-rzCT}
		r = \sqrt{\frac{1-z^2}{1+\vk^2\,z^2}}~,\qquad 
		\pi_r = \frac{\p z}{\p r} \pi_z = \frac{-\sqrt{1-z^2}(1+\vk^2\,z^2)^{3/2}}{z(1+\vk^2)} \pi_z~,
	\ee
	which is part of the $\Integers_2$-symmetry observed for $(\AdS_3 \times \Sph^3)_\eta$ \cite{Hoare:2014pna}. By this, the Hamiltonian reads
	\be
		\label{eq:FS-ham} 
		H = 
			\frac{(1-r^2)(1 + \vk^2 r^2)}{2}\pi_r^2
			+ \frac{\pi_\phi^2}{2\,r^2}~.
	\ee
	Generally, the coordinate $r$ parametrizes only half the FS, see Figure \ref{fig:FSplot}, but as the above Hamiltonian offers a simplified angular part, we prefer to work with \eqref{eq:FS-ham} over \eqref{eq:FS-hamz}. 
	
	We can split the Hamiltonian into an undeformed part $H_\as$ and a deformation $H_{\d\as}$,
	\be
		\label{eq:FS-hamSdS}
		H = H_\as + \vk^2 H_{\d\as}~,\qquad
			H_\as = 
			\frac{1-r^2}{2}\pi_r^2
			+ \frac{\pi_\phi^2}{2\,r^2}~,\qquad 
			H_{\d\as} = 
			\frac{r^2(1-r^2)}{2}\pi_r^2~.
	\ee
	Hence, in the undeformed limit, $\vk \rightarrow 0$, the system reduces to $H_\as$, which describes geodesic motion on the two-sphere. Corresponding $\Reals^3$ embedding coordinates parametrizing half the $\Sph^2$ are
	\be
		\label{eq:FS-S2Embedd}
		\vec{X}_\as = \left(X^1_\as,X^2_\as,X^3_\as\right) = \left(r\cos(\phi), r\sin(\phi), \sqrt{1-r^2}\right)\qquad\text{with}\quad 0\leq r \leq 1~,
	\ee
	that is $\vec{X}_\as{}^2 = X^A_\as \d_{AB} X^B_\as =1$, for $A,B,\ldots = 1,2,3$ and $\d^{AB}=\d_{AB}=\diag(1,1,1)$ the $\Reals^3$ metric. With $\eps_{ABC}$ the Levi-Civita tensor, $\eps_{123}=1$, and the embedding momenta $P_{\as,A}= \d_{AB} \dot{X}^B_\as$, the $\SO(3)$ isometry group is generated by the angular momenta $J_{\as,A} = \eps_{ABC} \d^{BD} P_{\as,D} X_\as^C$,
	\be\label{eq:FS-S2Isom}
		\vec{J}_\as = \bigg(\frac{\sqrt{1-r^2}}{r}\big(\pi_\phi\cos(\phi) + r \pi_r \sin(\phi) \big), \frac{\sqrt{1-r^2}}{r} \big(\pi_\phi \sin(\phi) - r\pi_r \cos(\phi)\big), -\pi_\phi \bigg).
	\ee
	They fulfill the $\so(3)$ Poisson algebra and their quadratic Casimir is just the Hamiltonian $H_\as$,
	\be
		\label{eq:FS-so3alg}
		\{J_{\as,A},J_{\as,B}\}_{\rm PB} = \eps_{ABC} \d^{CD} J_{\as,D}~,\qquad \qquad
			H_\as =  \frac{1}{2} \vec{P}_\as{}^2 = \frac{1}{2} \vec{J}_\as{}^2~.
	\ee

	To investigate the $\vk\rightarrow\infty$ limit, let us apply the canonical rescaling
	\be
		\label{eq:FS-r2v}
		r = \vk^{-1}v~,\qquad\qquad \pi_r = \vk\,\pi_v~,
	\ee by which the Hamiltonian \eqref{eq:FS-ham} takes the form
	\be
		\label{eq:FS-hamHdH}
		H = \vk^2 \left(H_\ah + \vk^{-2} H_{\d\ah}\right)~,\qquad H_\ah = \frac{1+v^2}{2} \pi_v^2 +\frac{\pi_\phi^2}{2\,v^2}~,\qquad H_{\d\ah} = -\frac{v^2(1+v^2)}{2}\pi_v^2~.
	\ee
	Hence, in the $\vk \rightarrow\infty$ limit, the system is dominated by the Hamiltonian $H_\ah$, which describes geodesic motion on $d=2$ dimensional hyperbolic space alias Euclidean AdS, $\HS^2=\EAdS_2$. Here, the prefactor of $\vk^2$ could be absorbed into redefinition of the string tension $T$, see \eqref{eq:FS-sigmaLag}.

	Corresponding $\Reals^{1,2}$ embedding coordinates are
	\be
		\label{eq:FS-H2Embedd}
		\vec{X}_\ah = \left(X^1_\ah,X^2_\ah,X^3_\ah\right) = \left(v\cos(\phi), v\sin(\phi), \sqrt{1+v^2}\right)\qquad\text{with}\quad 0\leq v \leq \infty~,
	\ee
	that is $\vec{X}_\ah{}^2 = X^A_\ah \eta_{AB} X^B_\ah =-1$, for again $A,B,\ldots = 1,2,3$ but now $\eta^{AB}=\eta_{AB}=\diag(1,1,-1)$ the $\Reals^{1,2}$ metric. Analogously to \eqref{eq:FS-S2Isom}, with the embedding momenta $P_{\ah,A}= \eta_{AB} \dot{X}^B_\ah$, the $\SO(1,2)$ isometry group is generated by the boosts and angular momentum $J_{\ah,A} = \eps_{ABC} \eta^{BD} P_{\ah,D} X_\ah^C$,
	\be\label{eq:FS-H2Isom}
		\vec{J}_\ah = \bigg(\frac{\sqrt{1+v^2}}{v}\big(\pi_\phi\cos(\phi) + v \pi_v \sin(\phi) \big), \frac{\sqrt{1+v^2}}{v} \big(\pi_\phi \sin(\phi) - v\pi_v \cos(\phi)\big), -\pi_\phi \bigg).
	\ee
	They fulfill the $\so(1,2)$ Poisson algebra and their quadratic Casimir is just the Hamiltonian $H_\ah$,
	\be
		\label{eq:FS-so12alg}
		\{J_{\ah,A},J_{\ah,B}\}_{\rm PB} = \eps_{ABC} \eta^{CD} J_{\ah,D}~,\qquad \qquad
			H_\ah =  \frac{1}{2} \vec{P}_\ah{}^2 = \frac{1}{2} \vec{J}_\ah{}^2~.
	\ee
	Apart from noting the similarity between \eqref{eq:FS-S2Isom} and \eqref{eq:FS-H2Isom}, let us stress that both limiting cases possess the maximum amount of $2\,d-1=3$  non-abelian integrals of motion, {\it i.e.}, they are both maximally superintegrable.

	Of course the above limits are expected, as the sigma model on $(\AdS_2 \times \Sph^2)_\eta$ supposedly limits to the ones on $\AdS_2 \times \Sph^2$ and ${\rm dS}_2 \times \HS^2$ \cite{Hoare:2014pna}, respectively, which in turn could be viewed as a truncation of the limiting behaviour of $(\AdS_5 \times \Sph^5)_\eta$ \cite{Delduc:2013qra,Delduc:2014kha}.

	However, as exhibited in \cite{Arutyunov:2014cra}, there is another interesting candidate for the $\vk\rightarrow\infty$ limit of $(\AdS_5 \times \Sph^5)_\eta$ given by the so called mirror geometry of $\AdS_5 \times \Sph^5$, which we denote as $(\AdS_5 \times \Sph^5)_\alg{m}$. It is obtained by double Wick-rotating the world-sheet theory of the light cone $\AdS_5 \times \Sph^5$ string theory \cite{Arutynov:2014ota, Arutyunov:2014jfa}, see also \cite{Pachol:2015mfa}.
	For the $\AdS_5$-time $t$ and an $\Sph^5$-angle $\vp$ forming the light-cone coordinates, this effectively amounts to interchanging the metric components $g_{tt}$ and $1/g_{\vp\vp}$ as well as swapping the sign of the $B$-field.

	The hyperboloid $\HS^2=\EAdS_2$ discussed above could be viewed as a subspace of $(\AdS_5 \times \Sph^5)_\alg{m}$, but it is not $(\Sph^2)_\alg{m}$, the sphere part of $(\AdS_2 \times \Sph^2)_\alg{m}$, {\it i.e.}, of the mirror geometry obtained for $\AdS_2 \times {\rm S}^2$.\footnote{Although ${\rm dS}_2 \times \HS^2$ and $(\AdS_2 \times {\rm S}^2)_\alg{m}$ are related via a double T-duality \cite{Arutyunov:2014cra}.} The limit corresponding to the latter is obtained \cite{Arutyunov:2014cra} by the rescaling
	\be
		\label{eq:FS-wzCM}
		w = \vk\,z = \vk \sqrt{\frac{1-r^2}{1+\vk^2r^2}}~,\qquad\qquad \vp = \vk\,\phi~.
	\ee
	By this the Hamiltonian \eqref{eq:FS-hamz} becomes
	\be
		\label{eq:FS-hamMdM} 
		H = \vk^2 \left(H_\alg{m} + \ord(\vk^{-2})\right)\,,\qquad \quad
			H_\alg{m} = \frac{1+w^2}{2} \left(\pi_w^2 + \pi_\vp^2\right)\,,
	\ee
	where one can read off the metric of $({\rm S}^2)_\alg{m}$, $\dd s^2 = \frac{1}{1+w^2}\left(\dd w^2 + \dd \vp^2\right)$.

	In the present work, we are mainly interested in the isometry algebras for $\Sph^2$ \eqref{eq:FS-so3alg} and $\HS^2$ \eqref{eq:FS-so12alg} and the resulting maximal superintegrability of the geodesic motion. As a byproduct, our analysis will yield results for the geodesic motion on $({\rm S}^2)_\alg{m}$.

	With the above limiting cases, let us try to construct embedding coordinates for the FS corresponding to \eqref{eq:FS-ham}. But immediately one has to wonder what the signature of this embedding space should be, whether it should be $\Reals^3$ or $\Reals^{1,2}$. Taking for definiteness $\Reals^3$ and the ansatz
	\be
		\label{eq:FS-FSembedd}
		\vec{X} = \big(r \cos(\phi), r \sin(\phi), Z(r) \big)\,,
	\ee
	we find $\dd s^2 = \dd \vec{X}^2 = \frac{\dd r^2}{(1-r^2)(1+\vk^2\,r^2)} + r^2 \dd \phi^2$ for
	\be
		\label{eq:FS-FSembedd2}
		Z(r)=\frac{1}{\vk} \ellE\left(\arctan(\vk\,z)\big|1+\vk^2\right)+Z_0~,
	\ee
	for $\ellE(\Psi|\mm)$ the incomplete elliptic integral of the second kind, $m$ its modulus squared\footnote{We use the notation $\ellE(\Psi|\mm)=\int^\Psi_0 \dd\th\sqrt{1-\mm \sin^2(\th)}\,$.} and $z = \sqrt{\frac{1-r^2}{1+\vk^2\,r^2}}$ given in \eqref{eq:FS-rzCT}. Setting the integration constant $Z_0 = 0$, by $\p_r Z(r) \propto \sqrt{1-\vk^4 z^2}$ 
we have
	\be\ba
		\label{eq:FS-ZrBehavior}
		&Z(r) \in \Reals~, &&\text{for}&& \vk^2\,z \leq 1~,\\
		&Z(r) \in i\,\Reals + \Re(Z(\vk^{-2}))~,\qquad &&\text{for}\quad&& \vk^2\,z \geq 1~.
	\ea\ee

	This comes to no surprise, as it just reflects the fact that for $\vk^2\,z \leq 1$ and for $\vk^2\,z \geq 1$ the appropriate embedding spaces are $\Reals^3$ and $\Reals^{1,2}$, respectively. Put differently, such problems can be circumvented by embedding the FS into $\Reals^{1,3}$, with metric $\eta_{MN}=\diag(1,1,1,-1)$ and coordinates
	\be
		\vec{\mathbb{X}} = \big(r \cos(\phi),r \sin(\phi), \Re(Z(r)), \Im(Z(r))\big)~.
	\ee

	But as these are difficult to visualize, in the following we will instead plot the three-dimensional coordinates
	\be
		\label{eq:FS-FSembedd3}
		\vec{X} = \left(r \cos(\phi), r \sin(\phi), \Re(Z(r))-\Im(Z(r))\right)~. 
	\ee
	Indeed, $X^3 = \Re(Z(r))-\Im(Z(r))$ is not only continuous but also has a continuous first derivative. The corresponding $d=2$ dimensional FS manifolds are plotted in Figure \ref{fig:FSplot}.
	\begin{figure}[ht]
		\centering
		\begin{minipage}[b]{0.31\linewidth}
		\includegraphics[width=5cm]{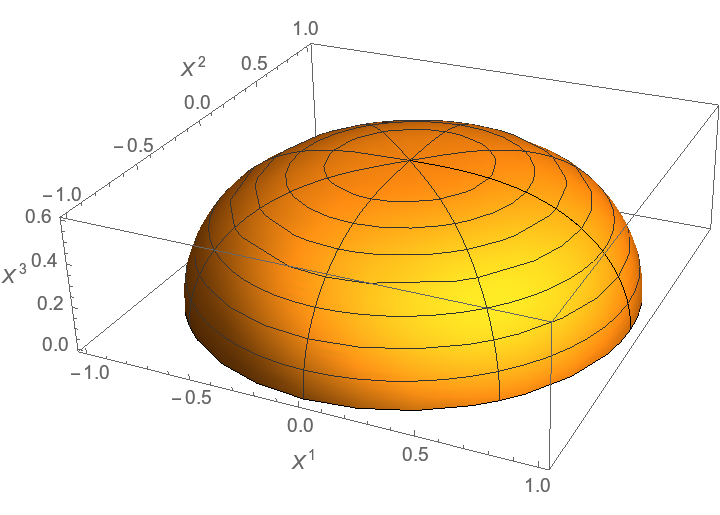}
		\end{minipage}
		\quad
		\begin{minipage}[b]{0.31\linewidth}
		\includegraphics[width=5cm]{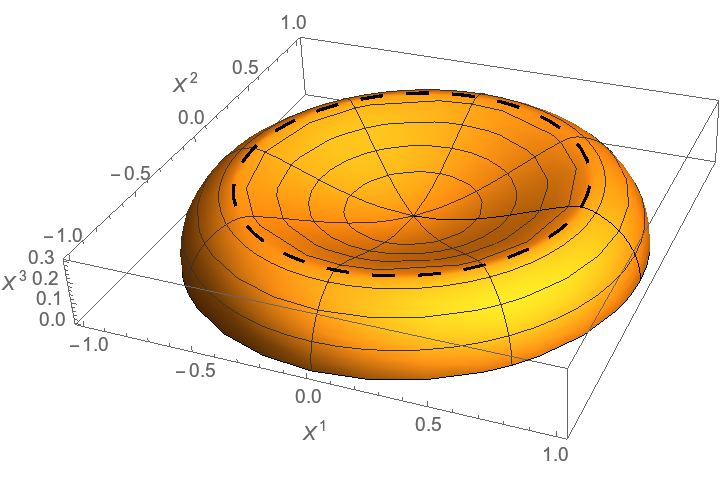}
		\end{minipage}
		\quad
		\begin{minipage}[b]{0.31\linewidth}
		\includegraphics[width=5cm]{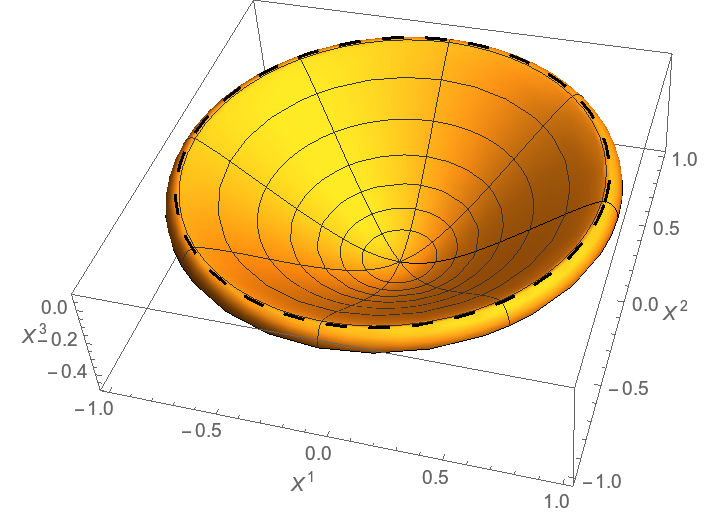}
		\end{minipage}
		\caption{Embedding of the FS, with $X^3 = \Re(Z(r))-\Im(Z(r))$, for $\vk=1$, $\vk=1.5$, and $\vk=3$.}
		\label{fig:FSplot}
	\end{figure}

	The left plot, for $\vk=1$, was shown previously in \cite{Arutyunov:2015qva}. Then for $\vk>1$ there appears a region $\vk^2 z\geq1$, where the shape of the FS becomes concave. Already for $\vk=3$ we see that the FS looks a lot like zoomed-out $d=2$ dimensional hyperbolic space ${\rm H}^2 = {\rm EAdS}_2$, a behaviour which gets more and more pronounced with increasing $\vk$. 

	By \eqref{eq:FS-ZrBehavior}, $z=\vk^{-2}$ corresponds to the seam ${\rm Max}(X^3)$ of the FS, plotted as dashed black line, which therefore marks the ``transition'' between Euclidean $\Reals^3$ and Lorentzian $\Reals^{1,2}$ signature.

\section{New integrals of motion}\label{sec:PIFS}

	From the embedding coordinates for the FS, \eqref{eq:FS-FSembedd} with \eqref{eq:FS-FSembedd2}, see also Figure \ref{fig:FSplot}, it is clear that for non-vanishing deformation, $\vk>0$, the two $\Reals^3$-isometries \eqref{eq:FS-S2Isom} $J_{\as,1}$ and $J_{\as,2}$ are broken, leaving only $J_{\as,3}=-\pi_\phi$ intact. Of course, this observation just reflects the fact, that general $\eta$-deformations leave only the corresponding Cartan subalgebra undeformed.

	We can however ask a distinct question: Apart from $J_{\as,3}=-\pi_\phi$, does the geodesic motion on the FS \eqref{eq:FS-ham} have additional integrals of motion? 

	To motivate this question, we recall that the limiting cases, the geodesic problem on $\Sph^2$ for $\vk\rightarrow0$ and on $\HS^2$ for $\vk\rightarrow\infty$, are maximally superintegrable. The corresponding non-abelian integrals of motion were given in \eqref{eq:FS-S2Isom} and \eqref{eq:FS-H2Isom}, respectively, and we are therefore wondering what happens to them ``in between'' these limits, that is for finite $\vk$.

\subsection{Constructing new integrals of motion}
	For definiteness, let us concentrate on the undeformed $\vk\rightarrow0$ limit. Having the Hamiltonian in the form $H=H_\as+\vk^2 H_{\d\as}$ \eqref{eq:FS-hamSdS}, at least for $\vk^2$ small one could perceived the term $\vk^2 H_{\d\as}$ as a perturbation of the geodesic motion on the two-sphere. Hence, we can try to construct integrals of motions $K_A$ by taking a power series ansatz in $\vk^2$ and demanding $K_A$ to limit to $J_{\as,A}$ as $\vk\rightarrow0$,
	\be
		\label{eq:FS-Kans}
		K_{A=1,2,3} = \sum_{n=0}^\infty \vk^{2n} K^{(n)}_{A} = J_{\alg{s},A} + \sum_{n=1}^\infty \vk^{2n} K^{(n)}_{A} 
			= J_{\alg{s},A} + \vk^{2} K^{(1)}_{A} + \ord(\vk^4)~.
	\ee
	Imposing $\{H,K_A\}_{\rm PB}=0$ to all orders in $\vk^2$ then yields at order  $\vk^{2n}$
	\be
		\label{eq:FS-KPDE}
		\ord\left(\vk^{2n}\right):\qquad\qquad \{H_{\d\as},K^{(n-1)}_A\}_{\rm PB} + \{H_{\alg{s}},K^{(n)}_A\}_{\rm PB} = 0~.\qquad\qquad
	\ee
	Assuming that we have solved \eqref{eq:FS-KPDE} to $\ord\left(\vk^{2(n-1)}\right)$, the order $\ord\left(\vk^{2n}\right)$ yields a partial differential equation for $K^{(n)}_A$ with inhomogeneity $\{H_{\d\as},K^{(n-1)}_A\}_{\rm PB}$. 

	As $\pi_\phi$ is already an integral of motion, we have
	\be
		\label{eq:FS-K3}
		K_3 = J_{\as,3} = J_{\ah,3} =-\pi_\phi~,
	\ee
	{\it i.e.}, set $K^{(n\geq1)}_3=0$, and we are left with finding $K_1$ and $K_2$. At first order in $\vk^2$ this reads
	\be
		\{H_{\d\as},J_{\alg{s},A=1,2}\}_{\rm PB} + \{H_{\alg{s}},K^{(1)}_{A=1,2}\}_{\rm PB} = 0~.
	\ee
	From \eqref{eq:FS-S2Isom} we see that $J_{\alg{s},A=1,2}$ has only $\cos(\phi)$- and $\sin(\phi)$-terms. Let us also expand $K^{(1)}_{A=1,2}$ in $\cos(l \phi)$- and $\sin(l \phi)$-modes,
	\be
		K^{(1)}_A(r,\phi,\pi_r,\pi_\phi)= K^{(1)}_{A,0}(r,\pi_r,\pi_\phi) + \sum^\infty_{l=1} \cos(l\phi) K^{(1)}_{A,c^l}(r,\pi_r,\pi_\phi) +\sin(l\phi) K^{(1)}_{A,s^l}(r,\pi_r,\pi_\phi)~.
	\ee
	Due to $\phi$-independence of $H_{\d\as}$ and $H_\alg{s}$, we observe that $\cos(l \phi)$ and $\sin(l \phi)$-modes for different $l$ decouple from each other. Especially, the only modes coupling to the inhomogeneity $\{H_{\d\as},J_{\alg{s},A}\}_{\rm PB}$ are the $l=1$ modes, $K^{(1)}_{A,c}$ and $K^{(1)}_{A,s}$. But by induction, the same holds to any order, {\it i.e.}, we can set all $\cos(l \phi)$ and $\sin(l \phi)$-modes for $l>1$ as well as the zero-modes to zero,
	\be
		\label{eq:FS-K12}
		K^{(n)}_{A=1,2}(r,\phi,\pi_r,\pi_\phi) = \cos(\phi) K^{(n)}_{A,c}(r,\pi_r,\pi_\phi) + \sin(\phi) K^{(n)}_{A,s}(r,\pi_r,\pi_\phi)~.
	\ee

	Next, we note that $H_{\alg{s}}$ as well as $H_{\d\as}$ have homogeneous power in momenta, they are both quadratic in $\{\pi_r,\pi_\phi\}$. For $\{H_{\alg{s}},K^{(1)}_{A=1,2}\}_{\rm PB}$ to cancel $\{H_{\d\as},J_{\alg{s},A}\}_{\rm PB}$, it follows that $K^{(1)}_{A=1,2}$ has to have the same powers in momenta as $J_{\alg{s},A}$. Especially, as $J_{\alg{s},A}$ is linear in momenta \eqref{eq:FS-S2Isom} so is $K^{(1)}_{A=1,2}$. But by induction, the same has to hold for all $K^{(n)}_{A=1,2}$, giving
	\be
		\label{eq:FS-K12-2}
		K^{(n)}_{A=1,2}(r,\phi,\pi_r,\pi_\phi) = \pi_\phi \left(\cos(\phi) f^{(n)}_{A,c}\left(r,\frac{\pi_r}{\pi_\phi}\right) + \sin(\phi) f^{(n)}_{A,s}\left(r,\frac{\pi_r}{\pi_\phi}\right)\right)~,
	\ee
	with the functions $f^{(n)}_{A,s}\left(r,\frac{\pi_r}{\pi_\phi}\right)$ of power zero in momenta. By resumming, our ansatz becomes
	\be
		\label{eq:FS-K12-3}
		K_{A=1,2}(r,\phi,\pi_r,\pi_\phi) = \pi_\phi \left(\cos(\phi) f_{A,c}\left(r,\frac{\pi_r}{\pi_\phi}\right) + \sin(\phi) f_{A,s}\left(r,\frac{\pi_r}{\pi_\phi}\right)\right)~,
	\ee
	where now the dependence on the deformation parameter $\vk$ is implicit and, formally, we do not rely on $K_{A=1,2}$ having power series expansion in $\vk^2$ \eqref{eq:FS-Kans} anymore.

	Let us introduce the following new variables of zero power in momenta,
	\be
		\label{eq:FS-PPkkdef}
		\PP = r\sqrt{1-r^2}\,\frac{\pi_r}{\pi_\phi}~,\qquad\qquad
		\kk   = \sqrt\frac{{2 H- \pi_\phi^2}}{\pi_\phi^2} = \sqrt{1-r^2}\sqrt{(1 + \vk^2 r^2) \frac{\pi_r^2}{\pi_\phi^2}
			+ \frac{1}{r^2}}~,
	\ee
	with the inverse transformation given by
	\be
		\label{FS-PPKKinv}
		r = \sqrt{\frac{1+\PP^2}{1+\kk^2-\vk^2 \PP^2}}~,\qquad
		\frac{\pi_r}{\pi_\phi} = \frac{(1+\kk^2-\vk^2 \PP^2) \PP}{\sqrt{1+\PP^2}\sqrt{\kk^2-(1+\vk^2)\PP^2}}~.
	\ee
	Note in particular that by $r\in[0,1]$ we have the inequality
	\be
		\label{eq:FS-ineq}
		\kk^2 = (1-r^2)\left((1 + \vk^2 r^2) \frac{\pi_r^2}{\pi_\phi^2}
			+ \frac{1}{r^2}\right) > (1-r^2)(1+\vk^2)r^2 \frac{\pi_r^2}{\pi_\phi^2} = (1+\vk^2) \PP^2~,
	\ee
	also implying $1+\kk^2 > \vk^2 \PP^2$. In terms of $\kk$ and $\PP$ the ansatz becomes
	\be
		\label{eq:FS-K12PPKK}
		K_{A=1,2}(r,\phi,\pi_r,\pi_\phi) = \pi_\phi \big(\cos(\phi) f_{A,c}\left(\kk,\PP\right) + \sin(\phi) f_{A,s}\left(\kk,\PP\right)\big)~.
	\ee

	The advantage of the variable $\kk$ defined in \eqref{eq:FS-PPkkdef} is that it manifestly is an integral of motion, $\{H,\kk\}_{\rm PB}=0$. Hence, after mode expansion in $\phi$, the defining equation $\{H,K_{A=1,2}\}_{\rm PB}=0$ yields only a coupled system of {\it ordinary} differential equations in $\PP$ instead of {\it partial} differential equations in the variable $r$ and $\pi_r/\pi_\phi$. Indeed, evaluating $\{H,K_{A=1,2}\}_{\rm PB}=0$ with \eqref{eq:FS-K12PPKK} we find
	\begin{align}
		\label{eq:FS-ffcsODEs}
		&f_{A,c}(\kk,\PP) = C_\beta(\kk,\PP)\,\frac{\p f_{A,s}(\kk,\PP)}{\p \PP}~,\qquad
			-f_{A,s}(\kk,\PP) = C_\beta(\kk,\PP)\,\frac{\p f_{A,c}(\kk,\PP)}{\p \PP}~,\\
		\label{eq:FS-Cbeta}
		&\qquad\qquad\qquad\qquad C_\beta(\kk,\PP) = \frac{(1+\PP^2)\sqrt{\kk^2 - (1+\vk^2) \PP^2}}{\sqrt{1+\kk^2 - \vk^2 \PP^2}}~,
	\end{align}
	with $C_\beta > 0$ due to \eqref{eq:FS-ineq}.
	
	To solve \eqref{eq:FS-ffcsODEs} we use the method of characteristics. That is, we want to find another variable $\beta(\kk,\PP)$, such that
	\be
		\label{eq:FS-betaODE}
		\frac{\p \beta}{\p \PP} = C_\beta(\kk,\PP)^{-1}~.
	\ee
	Plugging in \eqref{eq:FS-Cbeta}, this corresponds to an elliptic integral with the solution given by 
	\be\ba
		\label{eq:FS-betaSol}
		\beta(\kk,\PP) =& 
			\frac{1+\kk^2+\vk^2}{\sqrt{1+\vk^2}\sqrt{1+\kk^2}}\,\ellPi\left(\cc;\Phi\big|\mm\right) 
		- \frac{\vk^2}{\sqrt{1+\vk^2}\sqrt{1+\kk^2}}\,\ellF\left(\Phi\big|\mm\right) + g_\beta(\kk)~.
	\ea\ee
	Here, $\ellF\left(\Phi \big| \mm\right)$ and $\Pi\left(\cc;\Phi \big| \mm\right)$ are incomplete elliptic integrals of the first and third kind, respectively, having the argument $\Phi$, modulus squared $\mm$, and characteristic $\cc$,
	\begin{align}
		\label{eq:FS-EllArgs}
		&\Phi = \arcsin\left(\sqrt{\frac{1+\vk^2}{\kk^2}}\PP\right)\,,\qquad
		\mm = \frac{\vk^2 \kk^2}{(1+\vk^2)(1 + \kk^2)}~,\qquad
			\cc = \frac{-\kk^2}{1+\vk^2}~.
	\end{align}
	The function $g_\beta(\kk)$ corresponds to the homogeneous solution of \eqref{eq:FS-betaODE} and is, at this point, still undetermined.

	Having found the variable $\beta(\kk,\PP)$ fulfilling \eqref{eq:FS-betaODE} the ODE's \eqref{eq:FS-ffcsODEs} become\footnote{In an abuse of notation we identify $f(\kk,\beta(\kk,\PP))$ with $f(\kk,\beta)$.}
	\be
		f_{A,c}(\kk,\beta) =  \p_{\beta} f_{A,s}(\kk,\beta)~,\qquad
			-f_{A,s}(\kk,\beta) =  \p_{\beta} f_{A,c}(\kk,\beta)~,
	\ee
	which are just wave equations having the solutions
	\begin{align}
		\label{eq:FS-fAcs}
		&f_{A,c}(\kk,\beta) = g_{A,c}(\kk) \cos(\beta) + g_{A,s}(\kk) \sin(\beta)~,\\[.2em]
		&f_{A,s}(\kk,\beta) = g_{A,c}(\kk) \sin(\beta) - g_{A,s}(\kk) \cos(\beta) ~.
	\end{align}

	We now set
	\be
		\label{eq:FS-gAcs}
		g_{1,s}(\kk) = g_{2,c}(\kk) = 0~,\qquad\quad
		g(k) \equiv g_{1,c}(\kk) =
		-g_{2,s}(\kk) = \sqrt{\frac{\kk^2 + h(\vk)}{1 - h(\vk)}}~,
	\ee
	for a yet to be determined function $h(\vk)$ only depending on $\vk$.\footnote{From \eqref{eq:FS-hzero} on we set $h(\vk)=0\,$.} This gives us
	\be\ba
		\label{eq:FS-K12final}
		&K_1 = \pi_\phi\,g(\kk) \left(\cos{\phi}\cos{\beta}+\sin{\phi}\sin{\beta}\right)=\pi_\phi\,g(\kk) \cos(\phi-\beta)~,\\
		&K_2 = \pi_\phi\,g(\kk) \left(\sin{\phi}\cos{\beta}-\cos{\phi}\sin{\beta}\right) = \pi_\phi\,g(\kk) \sin(\phi-\beta)~.
	\ea\ee

\subsection{Undeformed limit to \texorpdfstring{$\Sph^2$}{S2}}
	
	With this, let us check the undeformed limit $\vk\rightarrow0$. The goal to reproduce \eqref{eq:FS-Kans}, $K_A \big|_{\vk=0} = J_{\as,A}$, now fixes the function $h(\vk)$ in this limit. In particular, requiring
	\be
		\label{eq:FS-hvkzero}
		h(\vk) \big|_{\vk=0} = 0~,
	\ee
	from \eqref{eq:FS-gAcs} and \eqref{eq:FS-PPkkdef} we have
	\be
		\label{eq:FS-jjsDef}
		g(\kk)\big|_{\vk=0} = \kk\big|_{\vk=0} 
		= \frac{\sqrt{1-r^2}}{r}\sqrt{\frac{r^2\pi_r^2}{\pi_\phi^2}+1} 
		= \sqrt{\frac{J_{\alg{s},1}^2+J_{\alg{s},2}^2}{J_{\alg{s},3}^2}} \equiv \jj_\as~.
	\ee
	For $\beta(\kk,\PP)$, let us assume for now that $g_\beta(\kk^2)\big|_{\vk=0} = 0$, where a discussion is postponed to \eqref{eq:NCC-galphabeta2}. Then, due to $\mm|_{\vk=0}=0$ and $\ellF(\Phi|0)=\Phi$, for $\beta(\kk,\PP)$ we get
	\be
		\label{eq:FS-betalim}
		\beta(\kk,\PP)|_{\vk=0} 
			= \sqrt{1+\jj_\as^2}\;\ellPi\left(\cc;\Phi\big|0\right)\!\Big|_{\vk=0} 
			= \arctan\left(\sqrt{\frac{1+\kk^2}{\kk^2-\PP^2}}\,\PP\right)\!\bigg|_{\vk=0}
			= \arctan\left(\frac{r\,\pi_r}{\pi_\phi}\right),
	\ee
	hence
	\be
		\pi_\phi \kk\,\sin(\beta)|_{\vk=0} = \frac{\sqrt{1-r^2}}{r} r\pi_r~,\qquad
			\pi_\phi \kk\,\cos(\beta)|_{\vk=0} = \frac{\sqrt{1-r^2}}{r} \pi_\phi~.
	\ee
	Therefore, in addition to $K_3 = J_{\as,3}$ \eqref{eq:FS-K3}, as intended \eqref{eq:FS-S2Isom} we have
	\be\ba
		&K_1|_{\vk=0} =\frac{\sqrt{1-r^2}}{r} \big(\pi_\phi\,\cos\phi +r\pi_r\,\sin\phi\big)=J_{\alg{s},1}~,\\
		&K_2|_{\vk=0} =\frac{\sqrt{1-r^2}}{r} \big(\pi_\phi\,\sin \phi - r\pi_r\, \cos \phi\big)=J_{\alg{s},2}~.
	\ea\ee

\subsection{Poisson algebra}
	
	Next, we would like to calculate the Poisson bracket algebra of $K_{A=1,2,3}$. But before doing so, recall that for the $\eta$-deformed $\AdS_5 \times \Sph^5$ sigma model, the original global superisometries $\psu(2,2|4)$ get replaced by a classical analog of the quantum group $\,\mathcal{U}_q\big(\psu(2,2|4)\big)$ \cite{Delduc:2014kha}, where $q=\exp\left(\frac{\vk}{\sqrt{1+\vk^2}}T \right)$ is the real quantum group parameter. Hence, it is reasonable to expect some truncation of this to happen in our system, {\it viz.}, that the $\so(3)$ symmetry \eqref{eq:FS-so3alg} is replaced by the quantum group $\,\mathcal{U}_q\big(\so(3)\big)$ and that $K_A$ fulfill the corresponding $q$-Poisson-Serre relations. With this expectation let us come back to the calculation.

	By $\{K_3,\cdot \}_{\rm PB} = \{-\pi_\phi,\cdot \}_{\rm PB} = -\p_\phi\,\cdot\,$ and $\p_\phi \kk = \p_\phi \PP = \p_\phi \beta =0$ we immediately have
	\be\ba
		\label{eq:FS-PBK3K12}
		&\{K_3,K_1\}_{\rm PB} = \pi_\phi\,g(\kk) \sin(\phi-\beta) = K_2~,\\
		&\{K_3,K_2\}_{\rm PB} = -\pi_\phi\,g(\kk) \cos(\phi-\beta) = -K_1~.
	\ea\ee
	Due to $\{\phi,\kk \}_{\rm PB} \neq 0$ and $\{\phi,\PP\}_{\rm PB} \neq 0$, calculation of $\{K_1,K_2\}_{\rm PB}$ seems much more difficult. Luckily, we can omit explicit calculation by noting that
	\be
		\label{eq:FS-calKsq}
		\frac{K_1{}^2 +K_2{}^2}{K_3{}^2} = g(\kk)^2 \Big(\cos^2(\phi-\beta)+\sin^2(\phi-\beta)\Big) = g(\kk)^2 = \frac{k^2 + h(\vk)}{1-h(\vk)}~.
	\ee
	Defining $\vec K = (K_1,K_2,K_3)$, we then calculate the $\so(3)$ scalar
	\be
		\label{eq:FS-Casimir}
		\frac{1}{2} \vec K^2 = \frac{1}{2} K_A \d^{A B} K_B
			= \frac{\pi_\phi^2}{2}\left(\frac{1+\kk^2}{1- h(\vk)}\right) 
			= \frac{H}{1- h(\vk)}~,
	\ee
	where we used the explicit expression for $\kk$ \eqref{eq:FS-PPkkdef}. Hence, up to the scaling factor $1-h(\vk)$, the Hamiltonian $H$ is nothing but the quadratic $\so(3)$ Casimir, which fits together with the fact that the $K_A$ are integrals of motion, $\{H, K_A\}_{\rm PB} = 0$, and which resembles the undeformed limit \eqref{eq:FS-so3alg}. But by the above we have
	\be
		0=\{K_1,H\}_{\rm PB} \propto K_2 \{K_1,K_2\}_{\rm PB} + K_3 \{K_1,K_3\}_{\rm PB} = K_2 \big(\{K_1,K_2\}_{\rm PB} - K_3\big)~,
	\ee
	where in the last step we plugged in \eqref{eq:FS-PBK3K12}, hence
	\be
		\label{eq:FS-PBK1K2}
		\{K_1,K_2\}_{\rm PB} = K_3~.
	\ee

  Combining \eqref{eq:FS-PBK1K2} with \eqref{eq:FS-PBK3K12} we conclude that
	\be
		\label{eq:FS-PBKK}
		\{K_A,K_B\}_{\rm PB} = \eps_{A B C} \d^{C D} K_D~.
	\ee
	That is, at arbitrary $\vk$ and for arbitrary $h(\vk)$ the new integrals of motion $K_A$ form the standard $\so(3)$ algebra. In particular, they do not fulfill some quantum deformed Poisson-Serre relations. Similarly, the $K_A$ do not seem to grant access to the Hopf-algebra structure of what should be a quantum deformed isometry algebra $\,\mathcal{U}_q\big(\so(3)\big)$.

	But maybe  one should not hope to witness such structure for the geodesic problem. After all, the $\,\mathcal{U}_q\big(\psu(2,2|4)\big)$ quantum algebra charges are a feature of the sigma model \cite{Delduc:2014kha}, with their derivation relying on non-trivial behavior along the spatial world-sheet direction.

	Then again, we should point out that the choice of the homogeneous solutions \eqref{eq:FS-gAcs} is tailored to give an $\so(3)$ algebra for the $K_A$. A different choice might allow to reproduce $q$-Poisson-Serre relations, 
	but just this as a guiding principle seems insufficiently motivated for geodesic motion.

\subsection{Infinitely deformed limit to \texorpdfstring{$\HS^2$}{HS2}}

	With such an immediate connection between the algebras of the $K_A$ and the one of the $\Sph^2$ isometries $J_{\as,A}$ \eqref{eq:FS-so3alg}, one has to wonder how the $\HS^2$ isometries $J_{\ah,A}$, which ought to correspond to the $\vk\rightarrow\infty$ limit of the $K_A$, can form an $\so(1,2)$ algebra \eqref{eq:FS-so12alg} instead. To reveal the relation, we employ once more the canonical rescaling \eqref{eq:FS-r2v}, $r=\vk^{-1} v$ and $\pi_r=\vk \,\pi_v$, yielding for $\PP$ and $\kk$ \eqref{eq:FS-PPkkdef},
	\be
		\PP = v \sqrt{1-\vk^{-2}v^2} \frac{\pi_v}{\pi_\phi}~,\qquad\qquad
			\kk = \vk \frac{\sqrt{1- \vk^{-2} v^2}}{\pi_\phi}\sqrt{(1+v^2)\pi_v^2 +\frac{\pi_\phi}{v}}~.
	\ee
	
	Hence, for $\vk\rightarrow\infty$ the modulus squared $m = \frac{\vk^2 \kk^2}{(1+\vk^2)(1+\kk^2)}$ \eqref{eq:FS-EllArgs} approaches $m=1$. Expanding the elliptic integrals in $\beta(\kk,\PP)$ at this point,
	\begin{align}
		&\ellPi(n;\Phi|m) = \frac{\sqrt{n}}{2(n-1)}\log\left(\frac{1+\sqrt{n} \sin(\Phi)}{1-\sqrt{n}\sin(\Phi)}\right) - \frac{\log\big(\sec(\Phi)+\tan(\Phi)\big)}{n-1} + \ord(m-1)~,\\
		&\ellF(\Phi|m) = \arctan\big(\sin(\Phi)\big) + \ord(m-1)
	\end{align}
	and assuming $g_\beta(k)\big|_{\vk\rightarrow\infty}=0$, for $\beta(\kk,\PP)$ we find
	\be
		\label{eq:FS-betaInf}
		\beta(\kk,\PP)\big|_{x\rightarrow\infty} = \frac{-i}{2} \log\left(\frac{\pi_\phi + i\,v\pi_v}{\pi_\phi - i\,v\pi_v}\right) = \arctan\left(\frac{v \pi_v}{\pi_\phi}\right)~.
	\ee
	
	The goal to obtain the $\so(1,2)$ algebra \eqref{eq:FS-so12alg} now fixes the behaviour of $h(\vk)$ in this limit. In particular, demanding
	\be
		\label{eq:FS-hvkinf}
		h(\vk) \big|_{\vk\rightarrow\infty} = \vk^2~
	\ee
	in analogy to \eqref{eq:FS-jjsDef} we have
	\be
		\label{eq:FS-jjhDef}
		g(k)\big|_{\vk\rightarrow\infty} = i \sqrt{1+\frac{k^2}{\vk^2}} \,\bigg|_{\vk\rightarrow\infty}
			= i \frac{\sqrt{1+v^2}}{v} \sqrt{\frac{v^2 \pi_v^2}{\pi_\phi^2} + 1}
			= i \sqrt{\frac{J_{\alg{h},1}^2+J_{\alg{h},2}^2}{J_{\alg{h},3}^2}} \equiv i \jj_\ah~, 
	\ee

	By this we have
	\be
		\pi_\phi g(\kk)\,\sin(\beta)|_{\vk\rightarrow\infty} = i \frac{\sqrt{1+v^2}}{v} v\pi_v~,\qquad
			\pi_\phi g(\kk)\,\cos(\beta)|_{\vk\rightarrow\infty} = i \frac{\sqrt{1+v^2}}{v} \pi_\phi
	\ee
	and therefore, apart from $K_3 = J_{\ah,3}$ \eqref{eq:FS-K3}, 
	\be\ba
		&K_1\big|_{\vk\rightarrow\infty} = i \frac{\sqrt{1+v^2}}{v}\big(\pi_\phi\cos(\phi) + v \pi_v \sin(\phi) \big) = i J_{\ah,1}~,\\
		&K_2\big|_{\vk\rightarrow\infty} = i \frac{\sqrt{1+v^2}}{v} \big(\pi_\phi \sin(\phi) - v\pi_v \cos(\phi)\big) = i J_{\ah,2}~.
	\ea\ee
	From this it is apparent that $\vec{K}|_{\vk\rightarrow\infty} = (i J_{\ah,1},i J_{\ah,2},J_{\ah,3})$ fulfilling the $\so(3)$ algebra \eqref{eq:FS-PBKK} is equivalent to $\vec{J}_{\ah}=( J_{\ah,1},J_{\ah,2},J_{\ah,3})$ fulfilling the $\so(1,2)$ algebra \eqref{eq:FS-so12alg}. But with $K_A$ taking complex values, it seems more appropriate to perceive \eqref{eq:FS-PBKK} as a complex Lie algebra. That is, the $K_A$ fulfill the complex $\alg{sl}(2)$ algebra with the limits \eqref{eq:FS-so3alg} and \eqref{eq:FS-so12alg} constituting particular reality conditions.

	A simple choice for $h(\vk)$ satisfying the limits \eqref{eq:FS-hvkzero} and \eqref{eq:FS-hvkinf} is given by
	\be
		h(\vk) = a \vk + \vk^2~.
	\ee
	Here, inclusion of the linear term in $\vk$ with arbitrary constant $a$ exhibits that there is no particular value of $\vk$, where $(1-h(\vk))^{-1}$ in \eqref{eq:FS-gAcs} and \eqref{eq:FS-Casimir} diverges, respectively, where $K_{1,2}$ switch from real to imaginary.

	Furthermore, we note that \eqref{eq:FS-Casimir} suggests that the quadratic Casimir $\frac{1}{2}\vec{K}$ is the appropriate Hamiltonian for geodesic motion. Especially, taking the scaling behaviour \eqref{eq:FS-hvkinf} seriously, for $\frac{1}{2}\vec{K}$ in the $\vk\rightarrow\infty$ limit there is no need to absorb a prefactor of $\vk^2$ into the string tension $T$ anymore, as promoted beneath \eqref{eq:FS-hamHdH}.

	Nevertheless, to simplify the following calculations, let us relax the requirement for $K_A$ to limit to $\vec{K}|_{\vk\rightarrow\infty} = (i J_{\ah,1},i J_{\ah,2},J_{\ah,3})$. In particular, for the rest of this work let us set
	\be
		\label{eq:FS-hzero}
		h(\vk) = 0~,
	\ee
	which still ensures the undeformed limit \eqref{eq:FS-Kans}, $\vec{K}|_{\vk=0} = \vec{J}_\as$, and keeps the $K_A$ real. Accordingly, for $g(k)$ and for the Casimir this yields
	\be
		\label{eq:FS-calKsq2}
		\sqrt{\frac{K_1{}^2 +K_2{}^2}{K_3{}^2}} = g(\kk) = \kk~,\qquad\qquad \frac{1}{2}\vec{K}^2 = \frac{\pi_\phi^2}{2}(1+k^2) = H~,
	\ee
	which resembles the limiting cases \eqref{eq:FS-jjsDef} and \eqref{eq:FS-jjhDef} and therefore elucidates our notation.

\subsection{Infinitely deformed limit to \texorpdfstring{$(\Sph^2)_\alg{m}$}{S2m}}
	From the integrals of motion $K_A$ on the FS we can now generate integrals of motion for geodesic motion on $(\Sph^2)_\alg{m}$, the mirror geometry of the two-sphere, described by the Hamiltonian \eqref{eq:FS-hamMdM}, $H_\alg{m} = \frac{1+w^2}{2}(\pi_w^2+\pi_\varphi^2)$. For this we employ the canonical maps \eqref{eq:FS-rzCT} and \eqref{eq:FS-wzCM} and then take the $\vk\rightarrow\infty$ limit. Setting $\PP_\alg{m} \equiv \vk\,\PP(w,\pi_w,\pi_\varphi)\big|_{\vk\rightarrow\infty}$ and $k_\alg{m} \equiv k(w,\pi_w,\pi_\varphi)\big|_{\vk\rightarrow\infty}$ this yields
	\be
		\PP_\alg{m} = -\sqrt{1+w^{-2}}\,\frac{w\,\pi_w}{\pi_\varphi} ~,\qquad\qquad
		\kk_\alg{m} 
			= \frac{w}{\pi_\varphi}\sqrt{(1+w^{-2})\pi_w^2 + \pi_\varphi^2}~.
	\ee
	
	As $K_{1,2}$ \eqref{eq:FS-K12final} depend on the angle $\phi-\beta$ whilst $\phi$ is subject to the canonical rescaling \eqref{eq:FS-wzCM}, $\phi = \vk^{-1} \varphi$, in the limit to $(\Sph^2)_\alg{m}$ we also need $\beta(w,\pi_w,\pi_\varphi) \big|_{\vk\rightarrow\infty} = \ord(\vk^{-1})$. This is indeed the case due to the scaling of the characteristic \eqref{eq:FS-EllArgs}, $n(w,\pi_w,\pi_\varphi) \big|_{\vk\rightarrow\infty} = \ord(\vk^{-1})$, and we find
	\be\ba
		\label{eq:FS-betam}
		&\beta_\alg{m} \equiv \vk\, \beta(w,\pi_w,\pi_\varphi) \big|_{\vk\rightarrow\infty} 
			= \sqrt{1+\kk_\alg{m}}\,\ellE\left(\arcsin\left(\frac{\PP_\alg{m}}{\sqrt{\kk_\alg{m}^2}}\right),\frac{\kk_\alg{m}}{1+\kk_\alg{m}^2}\right)\\
		=& -\sqrt{(1+w^2)\frac{\pi_w^2+\pi_\varphi^2}{\pi_\varphi^2}}\,\ellE\left(\arcsin\left(
				\frac{w\,\pi_w/\pi_\varphi}{\sqrt{\frac{w^4}{1+w^2}+ \frac{w^2\,\pi_w^2}{\pi_\varphi^2}}}\right),
			\frac{\pi_w^2+\frac{w^2}{1+w^2} \pi_\varphi^2}{\pi_w^2 +\pi_\varphi^2}  \right)~.
	\ea\ee

	Applying the limit directly to the FS integrals of motion, $K_A(w,\pi_w,\pi_\varphi) \big|_{\vk\rightarrow\infty}$, one gets
	\be
		K_1 \rightarrow \vk\,\pi_\varphi\,\kk_\alg{m}~,\qquad\quad 
			K_2 \rightarrow \pi_\varphi\,\kk_\alg{m} (\varphi -\beta_\alg{m})~,\qquad\quad
			K_3 \rightarrow \vk\,\pi_\varphi~.
	\ee
	Although $\frac{1}{2}\vec{K}^2 \rightarrow \vk^2\,H_\alg{m}$, this limit appears to be inconsistent as the $\so(3)$ algebra \eqref{eq:FS-PBKK} is spoiled.
	Therefore, in analogy to the form of the $K_A$ let us instead define
	\be
		J_{\alg{m},1}\equiv \pi_\varphi\,\kk_\alg{m}\,\cos(\varphi - \beta_\alg{m})~,\qquad
			J_{\alg{m},2}\equiv \pi_\varphi\,\kk_\alg{m}\,\sin(\varphi - \beta_\alg{m})~,\qquad
			J_{\alg{m},3}\equiv -\pi_\varphi~.
	\ee
	One can directly check that $\{H_\alg{m},J_{\alg{m},A}\}_{\rm PB}=0$. Furthermore, due to their resemblance to the $K_A$ the arguments leading to \eqref{eq:FS-PBKK} still apply, in particular $\frac{1}{2}\vec{J}_\alg{m}^{\,2} = \frac{1}{2} J_{\alg{m},A} \d^{AB} J_{\alg{m},B} = H_\alg{m}$ and the $J_{\alg{m},A}$ form an $\so(3)$ algebra.

\section{New canonical coordinates}\label{sec:NCC}
	The fact that the new integrals of motion $K_A$ form an undeformed $\so(3)$ algebra suggests that the geodesic motion on (half) the FS might be connected to geodesic motion on (half) the two sphere $\Sph^2$ via a highly non-trivial canonical transformation, {\it i.e.}, that there exist new phase space variables\footnote{Here, $\pit_r = \pi_\rt$ and $\pit_\phi = \pi_\phit$ are the momenta conjugate to $\rt$ and $\phit$, respectively.} $\{\rt,\phit,\pit_r,\pit_\phi\}$ such that expressed in these $H$ becomes
	\be
		\label{eq:FS-FSasS2}
		H = \frac{1-\rt^2}{2} \pit_r^2 + \frac{\pit_\phi^2}{2 \rt^2}
		\quad\qquad\text{with}\quad 0\leq \rt \leq1~.
	\ee
	When expressed in these phase space coordinates the integrals of motion $K_A$ have to take the form of \eqref{eq:FS-S2Isom} under the substitution $\{r,\phi,\pi_r,\pi_\phi\} \mapsto \{\rt,\phit,\pit_r,\pit_\phi\}$. In the following we construct the new canonical coordinates $\{\rt,\phit,\pit_r,\pit_\phi\}$ by brute force.

\subsection{Deriving \texorpdfstring{$\pit_\phi$, $\rt$, and $\pit_r$}{pitphi,rt,pitr}}
	First of all, by \eqref{eq:FS-K3} we take
	\be
		\label{eq:NCC-pitphi}
		\pit_\phi = -K_3 = \pi_\phi~.
	\ee
	Note that generally this does {\it not} imply $\phit = \phi\,$. However, from the fact that the undeformed limit of the FS is the two sphere $\Sph^2$, $H|_{\vk=0}=H_{\alg{s}}$, we know that
	\be
		\label{eq:NCC-rphi-vkzero}
		\rt = r +\ord(\vk^2)~,\qquad \pit_r = \pi_r + \ord(\vk^2)~,\qquad \phit = \phi + \ord(\vk^2)~.
	\ee
	By repeating the arguments in the previous section, the new coordinates $\rt$ and $\phit$ ought to have zero power, while $\pit_r$ is linear in old momenta $\pi_r$ and $\pi_\phi$. Using again the coordinates $\kk$ and $\PP$ \eqref{eq:FS-PPkkdef} this yields
	\be
		\label{eq:NCC-ansatz}
		\rt = \rt(\kk,\PP,\phi)~,\qquad \phit = \phit(\kk,\PP,\phi)~,\qquad
			\pit_r = \pi_\phi\,\varpi_r(\kk,\PP,\phi)~.
	\ee
	Imposing now $\{\pit_\phi,\rt\}_{\rm PB} = \{\pit_\phi,\pit_r\}_{\rm PB} = 0$ tells us that $\rt$ and $\pit_r$ do not depend on $\phi$,
	\be
		\label{eq:NCC-rtpirt}
		\rt = \rt(\kk,\PP)~,\qquad \pit_r = \pi_\phi \varpi_r(\kk,\PP)~.
	\ee

	Next, we can compare \eqref{eq:FS-calKsq2} with \eqref{eq:FS-FSasS2},
	\be
		2 H = \pi_\phi^2 \big(1+\kk^2\big) \stackrel{!}=  (1-\rt^2) \pit_r^2 + \frac{\pi_\phi^2}{\rt^2}
	\ee
	and solve for $\pit_r$,
	\be
		\label{eq:NCC-pirt}
		\pit_r(\kk,\PP) = \frac{\pm_\varpi \pi_\phi}{\rt(\kk,\PP)} \sqrt{\frac{(1+\kk^2)\rt(\kk,\PP)^2-1}{1-\rt(\kk,\PP)^2}}~,
	\ee
	where we assumed $\rt(\kk,\PP) \in [0,1]$ and the sign $\pm_\varpi$ is to be determined. This agrees with \eqref{eq:NCC-rtpirt}.

	Demanding $\rt$ and $\pit_r$ to be canonically conjugate, $\{\pit_r,\rt\}_{\rm PB} =1$, then results in 
	\begin{align}
		\label{eq:NCC-rtODE}
		&C_{\alpha}(\kk,\PP)\,\p_\PP \rt(\kk,\PP) = 
			\pm_\varpi \frac{\sqrt{1-\rt(\kk,\PP)^2}}{\sqrt{1+\kk^2}\,\rt(\kk,\PP)} \sqrt{\big(1+\kk^2\big)\rt(\kk,\PP)^2 -1}~,\\
		\label{eq:NCC-Calpha}
		&\qquad\qquad C_{\alpha} = \frac{\sqrt{1+\kk^2 -\vk^2 \PP^2}\sqrt{\kk^2 -(1+\vk^2) \PP^2}}{\sqrt{1+\kk^2}}~.
	\end{align}
	Especially, \eqref{eq:NCC-rtODE} is only an {\it ordinary} differential equation due to the use of the variable $\kk$ \eqref{eq:FS-PPkkdef}. To solve \eqref{eq:NCC-rtODE}, we again use the method of characteristics. In particular, similarly to \eqref{eq:FS-betaODE}, we are looking for a new variable $\alpha(\kk,\PP)$, such that
	\be
		\label{eq:NCC-alphaODE}
		\frac{\p \alpha}{\p \PP} = C_\alpha(\kk,\PP)^{-1}~.
	\ee
	Similarly to \eqref{eq:FS-betaODE}, this is an elliptic integral with the solution
	\be
		\label{eq:NCC-alphaSol}
		\alpha(\kk,\PP) =  \frac{1}{\sqrt{1+\vk^2}}\,\ellF\left(\Phi\big|\mm\right)  + g_\alpha(\kk)~,
	\ee
	with the argument $\Phi$ and modulus squared $\mm$ of the elliptic integral of the first kind given in \eqref{eq:FS-EllArgs},
	\be
		\label{eq:NCC-EllArgs2}
		\Phi = \arcsin\left(\sqrt{\frac{1+\vk^2}{\kk^2}}\PP\right)\,,\qquad
			\mm = \frac{\vk^2\kk^2}{(1+\vk^2)(1 + \kk^2)}~.
	\ee

	To gain a better appreciation for the new variable $\alpha(\kk,\PP)$, we note that
	\be
		\label{eq:NCC-KsqAlphaPB}
		\{\kk^2,\alpha\}_{\rm PB} 
			= \frac{2 \sqrt{1+\kk^2}}{\pi_\phi}~,\qquad\quad
			\text{sgn}(\pi_\phi)\left\{\abs{\vec{K}},\alpha\right\}_{\rm PB} = \pi_\phi\left\{\sqrt{1+\kk^2},\alpha\right\}_{\rm PB} = 1~,
	\ee
	{\it i.e.}, $\alpha$ is the angle conjugate to $\,\text{sgn}(\pi_\phi) \abs{\vec{K}}$ on the Liouville torus defined by $\abs{\vec{K}}$ and $\pi_\phi$.

	Having this new variable $\alpha(\kk,\PP)$, as anticipated the ODE for $\rt(\kk,\alpha)$ reads\footnote{Again, we abuse notation by identifying $\rt(\kk,\alpha)$ with $\rt(\kk,\alpha(\kk,\PP))$, etc.}
	\be
		\label{eq:NCC-rtODE2}
		\pm_\varpi \p_\alpha \rt(\kk,\alpha) = \frac{\sqrt{1-\rt(\kk,\alpha)^2}}{\rt(\kk,\alpha)} \sqrt{\rt(\kk,\alpha)^2 -\frac{1}{1+\kk^2}}~,
	\ee
	with the general solution
	\be
		\label{eq:NCC-rtSol}
		\rt(\kk,\alpha) = \frac{1}{4\sqrt{1+\kk^2}}\sqrt{e^{2i (\pm_\varpi \alpha + g_{\rt}(\kk))}+ 16\,\kk^4 e^{-2i (\pm_\varpi\alpha + g_{\rt}(\kk))}+ 8(2+\kk^2)}~.
	\ee
	The homogeneous solutions $g_{\alpha}(\kk)$ in \eqref{eq:NCC-alphaSol} and $g_{\rt}(\kk)$ in \eqref{eq:NCC-rtSol} just add up. We are therefore free to fix $g_{\rt}(\kk)$ arbitrarily without affecting the corresponding freedom. Especially, choosing
	\be
		g_{\rt}(\kk) = \frac{-i}{4} \ln(16\,\kk^4)
	\ee
	we get
	\be
		\label{eq:NCC-rtSol2}
		\rt(\kk,\alpha) 
			=\sqrt{\frac{1+\kk^2 \cos^2\left(\alpha\right)}{1+\kk^2}}~.
	\ee

	Note that the solution \eqref{eq:NCC-rtSol2} fulfills $0\leq \rt(\kk,\alpha)\leq 1$, as anticipated. A more thorough discussion of the values taken by $\rt(\kk,\alpha)$ is postponed to Subsection \ref{subsec:NCC-Range}.

	Let us now test the solution for $\rt(\kk,\alpha)$. By plugging in \eqref{eq:NCC-rtSol2} the ODE \eqref{eq:NCC-rtODE2} turns out to be proportional to
	\be
		\sqrt{\sin^2(2\alpha)} + \big(\pm_\varpi \sin(2\alpha)\big) \stackrel{!}= 0~,
	\ee
	which is fulfilled for
	\be \label{eq:NCC-alphaChoices}
		\ba
		&\pm_\varpi = -~,\qquad &&\alpha \in \left[0,\pi/2\right]+ s\,\pi~,\\
		\text{or}\quad &\pm_\varpi = +~,\qquad &&\alpha \in \left[-\pi/2,0\right]+ s \,\pi~,
	\ea\ee
	with $s\in\Integers$. From \eqref{eq:NCC-alphaSol} it is clear that the interval of $\alpha$ can be shifted arbitrarily by a constant contribution in $g_{\alpha}(\kk)$, such that both choices are equivalent. Especially, to fix notation, let us take the first choice of \eqref{eq:NCC-alphaChoices} with $s=0$,
	\be
		\label{eq:NCC-alphaInterval}
		\pm_\varpi \stackrel{!}= -\qquad \text{and} \qquad \alpha \stackrel{!}\in [0,\pi/2]~,
	\ee
	where we postpone further discussion to Subsection \ref{subsec:NCC-Range}.

 Plugging in \eqref{eq:NCC-rtSol2} into \eqref{eq:NCC-pirt} we then get 
	\be
		\label{eq:NCC-pitrSol}
		\pit_r(\kk,\alpha) =\pi_\phi \varpi_r(\kk,\alpha)= -\pi_\phi \frac{(1+\kk^2) \cot\left(\alpha\right)}{\sqrt{1+\kk^2 \cos^2\left(\alpha\right)}}~,
	\ee
	where we used $\sqrt{\cot^2{\alpha}}=\cot{\alpha}$ for $\alpha \in [0,\pi/2]$. With this we have
	\be
		\label{eq:NCC-rtsq-pitrrt}
		\rt^2 = \frac{1+\kk^2 \cos^2\left(\alpha\right)}{1+\kk^2}~,\qquad\qquad
		\rt\,\pit_r = -\pi_\phi \sqrt{1+\kk^2} \cot(\alpha)~,
	\ee
	and using \eqref{eq:NCC-KsqAlphaPB} we can check that $\rt$ and $\pit_r$ are indeed canonical,
	\be\ba
		\{\pit_r,\rt\}_{\rm PB} &= \frac{1}{2\,\rt^2}\{\rt\,\pit_r,\rt^2\}_{\rm PB} =  \frac{1}{2\,\rt^2} \{\kk^2,\alpha\}_{\rm PB}
			\left(\frac{\p (\rt\,\pit_r)}{\p \kk^2}\frac{\p\,\rt^2}{\p \alpha} -\frac{\p (\rt\,\pit_r)}{\p \alpha} \frac{\p\,\rt^2}{\p \kk^2}\right) =1~.
	\ea\ee

\subsection{Deriving \texorpdfstring{$\phit$}{phit}}\label{subsec:NCC-DerPhit}
	Now, we are left with finding $\phit(\phi,\kk,\PP) = \phi + \ord(\vk^2)$, cf. \eqref{eq:NCC-rphi-vkzero} and \eqref{eq:NCC-ansatz}, that is we are searching for a $\phit(\phi,\kk,\PP)$ such that
	\be
		\{\pit_\phi,\phit\}_{\rm PB} = \{\pi_\phi,\phit\}_{\rm PB} = 1~,\qquad
		\{\rt,\phit\}_{\rm PB} = 0~,\qquad
		\{\pit_r,\phit\}_{\rm PB} = 0~.
	\ee
	The first equation is easily solved by taking
	\be
		\label{eq:NCC-phitAnsatz}
		\phit(\phi,\kk,\PP) = \phi + \nu(\kk,\PP)~,
	\ee
	which is consistent with $\phit(\phi,\kk,\PP) = \phi + \ord(\vk^2)$ \eqref{eq:NCC-rphi-vkzero} for $\nu(\kk,\PP) =  \ord(\vk^2)$. 

	Equivalently to imposing the remaining Poisson brackets, let us consider the symplectic two form $\Omt$ of the new phase space variables. Using already known Poisson brackets one finds that
	\begin{align}
		\label{eq:NCC-twoForm}
		&\Omt = \dd \pit_r \wedge \dd \rt + \dd \pit_\phi \wedge \dd \phit
			= \dd \pi_r \wedge \dd r + \dd \pi_\phi \wedge \dd \phi
				+ \dd \pi_\phi \wedge \Big(\Omt_{\pi_\phi,r} \dd r + \Omt_{\pi_\phi,\pi_r} \dd \pi_r\Big)~,\\
		\label{eq:NCC-Omegas}
		\text{with}&\qquad \Omt_{\pi_\phi,r} 
			= \frac{\p \pit_r}{\p \pi_\phi}\frac{\p \rt}{\p r}
				-\frac{\p \pit_r}{\p r}\frac{\p \rt}{\p \pi_\phi} + \frac{\p \nu}{\p r}~,\qquad
			\Omt_{\pi_\phi,\pi_r} 
			= \frac{\p \pit_r}{\p \pi_\phi}\frac{\p \rt}{\p \pi_r}
				-\frac{\p \pit_r}{\p \pi_r}\frac{\p \rt}{\p \pi_\phi} + \frac{\p \nu}{\p \pi_r}~.
	\end{align}
	Hence, the transformation is canonical for $\Omt_{\pi_\phi,r}=\Omt_{\pi_\phi,\pi_r}=0$. Recalling that
	\be
		\frac{\p\rt}{\p \pi_\phi}=\frac{\p \rt\big(\!r,\frac{\pi_r}{\pi_\phi}\big)}{\p \pi_\phi} = \frac{-\pi_r}{\pi_\phi} \frac{\p\rt}{\p \pi_r}~,\qquad
		\frac{\p \pit_r}{\p \pi_\phi}=\frac{\p \pi_\phi\,\varpi_r\big(\!r,\frac{\pi_r}{\pi_\phi}\big)}{\p \pi_\phi} = \varpi_r\Big(\!r,\frac{\pi_r}{\pi_\phi}\Big) - ~ \frac{\pi_r}{\pi_\phi} \frac{\p \pit_r}{\p \pi_r},
	\ee
	and using once more $\{\pit_r,\rt\}_{\rm PB}=1$, \eqref{eq:NCC-Omegas} can be recast as
	\be
		\label{eq:NCC-Omegas2}
		\Omt_{\pi_\phi,r} 
			= \varpi_r \frac{\p \rt}{\p r}
				-\frac{\pi_r}{\pi_\phi} + \frac{\p \nu}{\p r} \stackrel{!}= 0~,\qquad\qquad
			\Omt_{\pi_\phi,\pi_r} 
			= \varpi_r \frac{\p \rt}{\p \pi_r} + \frac{\p \nu}{\p \pi_r} \stackrel{!}= 0~.
	\ee
	As expected, it can easily be shown that \eqref{eq:NCC-Omegas2} implies $\{\phit,\rt\}_{\rm PB} = \{\phit,\pit_r\}_{\rm PB} = 0$. 
	Furthermore, from this one can check that
	\be
		\frac{\p \Omt_{\pi_\phi,r}}{\p \pi_r} - \frac{\p \Omt_{\pi_\phi,\pi_r}}{\p r} = \frac{1}{\pi_\phi} \big(\{\pit_r,\rt\}_{\rm PB} -1 \big) = 0~,
	\ee
	which shows that the symplectic two-form $\Omt$ \eqref{eq:NCC-twoForm} is closed, $\dd \Omt=0$, for any $\Omt_{\pi_\phi,r}$ and $\Omt_{\pi_\phi,\pi_r}$ of the form \eqref{eq:NCC-Omegas2} but even without requiring them to vanish. Still, finding $\phit$ by imposing \eqref{eq:NCC-Omegas2} seems difficult, as the terms $\varpi_r \frac{\p \rt}{\p r}$ and $\varpi_r \frac{\p \rt}{\p \pi_r}$ will be highly complicated.

	Luckily, there is yet another way to obtain a candidate for $\phit\,$: If the Hamiltonian in terms of new phase space variables describes just the particle on $\Sph^2$, as anticipated in \eqref{eq:FS-FSasS2}, in these phase space coordinates the integrals of motion $K_A$, \eqref{eq:FS-K12final} and \eqref{eq:NCC-pitphi}, should just take the standard form of the isometries on $\Sph^2$ \eqref{eq:FS-S2Isom}, {\it i.e.}, they should read
	\be
		\label{eq:NCC-S2IsomK}
		\vec{K} = \pi_\phi\left(
			\frac{\sqrt{1-\rt^2}}{\rt}\big(\!\cos(\phit) + \sin(\phit)\,\rt\,\varpi_r\big),
			\frac{\sqrt{1-\rt^2}}{\rt} \big(\! \sin(\phit)-\cos(\phit)\,\rt\,\varpi_r\big),
			-1 \right)~.
	\ee
	Hence, comparing with \eqref{eq:FS-K12final} we find
	\be
		\frac{K_2}{K_1} = \frac{\tan{\phit}-\rt\,\varpi_r}{1 + \rt\varpi_r \tan{\phit}} = \tan(\phi-\beta)
	\ee
	and by using the arctangent addition formula
	\be
		\label{eq:NCC-phit}
		\phit = \arctan\left(\frac{\tan(\phi-\beta)+\rt\,\varpi_r}{1 - \rt\,\varpi_r \tan(\phi-\beta)}\right)
			= \phi - \beta + \arctan(\rt\,\varpi_r)~.
	\ee
	In particular, we see that the $\nu(\kk,\PP)$ of \eqref{eq:NCC-phitAnsatz} is given by
	\be
		\nu = \arctan(\rt\,\varpi_r) - \beta~.
	\ee
	
	What is left is to check \eqref{eq:NCC-Omegas2}. Amazingly, perceiving $\rt$ and $\varpi_r$ as functions of $\kk^2$ and $\alpha$, see \eqref{eq:NCC-rtSol2} and \eqref{eq:NCC-pitrSol}, we find
	\be
		\varpi_r\,\dd\rt + \dd \left(\arctan(\rt\,\varpi_r)\right)
			= \varpi_r\,\dd\rt +\frac{ \dd \left(\rt\,\varpi_r\right)}{1+(\rt\,\varpi_r)^2} =\sqrt{1+\kk^2}\,\dd \alpha~,
	\ee
	and the conditions \eqref{eq:NCC-Omegas2} become
	\be
		\label{eq:NCC-Omegas3}
		\Omt_{\pi_\phi,r} 
			= \sqrt{1+\kk^2}\,\frac{\p \alpha}{\p r}
				-\frac{\pi_r}{\pi_\phi} - \frac{\p \beta}{\p r} \stackrel{!}= 0~,\qquad
			\Omt_{\pi_\phi,\pi_r} 
			= \sqrt{1+\kk^2}\,\frac{\p \alpha}{\p \pi_r} - \frac{\p \beta}{\p \pi_r} \stackrel{!}= 0~.
	\ee
	Perceiving now $r$ and $\pi_r$ as functions of $\kk^2$, $\PP$, and $\pi_\phi$ we have
	\begin{align}
		\Omt_{\pi_\phi,r} \dd r + \Omt_{\pi_\phi,\pi_r} \dd \pi_r
			= \Omt_{\pi_\phi,\kk^2} \dd \kk^2 + \Omt_{\pi_\phi,\PP} \dd \PP + \Omt_{\pi_\phi,\pi_\phi} \dd \pi_\phi
	\end{align}
	with
	\be\ba
		\label{eq:NCC-Omegas4}
		&\Omt_{\pi_\phi,\kk^2} 
			= \sqrt{1+\kk^2}\,\frac{\p \alpha}{\p \kk^2}
				-\frac{\pi_r}{\pi_\phi}\,\frac{\p r(\kk^2,\PP)}{\p \kk^2} - \frac{\p \beta}{\p \kk^2} \stackrel{!}= 0~,\\
		&\Omt_{\pi_\phi,\PP} 
			= \sqrt{1+\kk^2}\,\frac{\p \alpha}{\p \PP}
				-\frac{\pi_r}{\pi_\phi}\,\frac{\p r(\kk^2,\PP)}{\p \PP} - \frac{\p \beta}{\p \PP} \stackrel{!}= 0~,
	\ea\ee
	and we ignore the $\Omt_{\pi_\phi,\pi_\phi}$ term, as it does not contribute in the symplectic two form, $\dd \pi_\phi \wedge \dd \pi_\phi=0$.

	From \eqref{eq:FS-betaODE} and \eqref{eq:NCC-alphaODE} we have $\p_\PP \beta(\kk^2,\PP) = C_{\beta}(\kk^2,\PP)^{-1}$ and $\p_\PP \alpha(\kk^2,\PP) = C_{\alpha}(\kk^2,\PP)^{-1}$.
	At the same time, by straight forward calculation we have
	\be
		\frac{\pi_r}{\pi_\phi}\frac{\p r(\kk^2,\PP)}{\p \PP}
			=  \frac{\sqrt{1+\kk^2}}{C_\alpha} - \frac{1}{C_\beta}~,
	\ee
	and therefore
	\be
		\Omt_{\pi_\phi,\PP} 
			= \frac{\sqrt{1+\kk^2}}{C_\alpha} - \left(\frac{\sqrt{1+\kk^2}}{C_\alpha} - \frac{1}{C_\beta}\right) - \frac{1}{C_\beta} = 0~.
	\ee

	Hence, we are left with showing that $\Omt_{\pi_\phi,\kk^2}=0$, for which we have to calculate $\p_{\kk^2} \alpha$ and $\p_{\kk^2} \beta$. Differentiating \eqref{eq:FS-betaSol} and \eqref{eq:NCC-alphaSol} we obtain
	\begin{align}
	 &\ba
		\frac{\p \beta(\kk^2,\PP)}{\p \kk^2} =&\,
			\frac{\sqrt{1+\kk^2}\sqrt{1+\vk^2}}{2\,\kk^2 (1+\kk^2 + \vk^2)} \ellE\left(\Phi,\mm\right) - \frac{\ellF\left(\Phi,\mm\right)}{2\,\kk^2 \sqrt{1+\kk^2}\sqrt{1+\vk^2}} \\
			&- \frac{(1+\vk^2)(1+\PP^2)\PP}{2\,\kk^2 (1+\kk^2+\vk^2) \,C_\beta} + \frac{\p g_\beta(\kk)}{\p\kk^2}~,\\
	\ea\\
	&\ba
		\frac{\p \alpha(\kk^2,\PP)}{\p \kk^2} =&\,
			\frac{\sqrt{1+\vk^2}}{2\,\kk^2 (1+\kk^2 + \vk^2)} \ellE\left(\Phi,\mm\right) - \frac{\ellF\left(\Phi,\mm\right)}{2\,\kk^2 (1+\kk^2)\sqrt{1+\vk^2}} \\
			&-\frac{\PP}{2(1+\kk^2)\,C_\alpha} - \frac{(1+\vk^2)(1+\PP^2)\PP}{2\sqrt{1+\kk^2}\kk^2(1+\kk^2+\vk^2) \,C_\beta} + \frac{\p g_\alpha(\kk)}{\p\kk^2}~,
	\ea
	\end{align}
	hence
	\be
		\sqrt{1+\kk^2}\,\frac{\p \alpha(\kk,\PP)}{\p \kk^2} - \frac{\p \beta(\kk,\PP)}{\p \kk^2} = \frac{-\PP}{2 \sqrt{1+\kk^2} C_\alpha} + \sqrt{1+\kk^2}\,\frac{\p g_{\alpha}(\kk)}{\p \kk^2} - \frac{\p g_{\beta}(\kk)}{\p \kk^2}~.
	\ee
	At the same time, we have
	\be
		\frac{\pi_r}{\pi_\phi}\frac{\p r(\kk^2,\PP)}{\p \kk^2}
			= \frac{- \PP}{2 \sqrt{1+\kk^2}\,C_\alpha}~
	\ee
	and therefore
	\be
		\Omt_{\pi_\phi,\kk^2} 
			= \frac{- \PP}{2 \sqrt{1+\kk^2}\,C_\alpha} + \sqrt{1+\kk^2}\,\frac{\p g_{\alpha}(\kk^2)}{\p \kk^2} - \left( \frac{- \PP}{2 \sqrt{1+\kk^2}\,C_\alpha}\right) - \frac{\p g_{\beta}(\kk^2)}{\p \kk^2}~,
	\ee
	which vanishes for
	\be
		\label{eq:NCC-galphabeta}
		\frac{\p g_{\beta}(\kk^2)}{\p \kk^2} = \sqrt{1+\kk^2}\,\frac{\p g_{\alpha}(\kk^2)}{\p \kk^2}~.
	\ee
	This shows, that under the condition \eqref{eq:NCC-galphabeta} the new phase space coordinates are indeed canonical. For the following discussion, let us therefore set
	\be
		\label{eq:NCC-galphabeta2}
		g_{\beta}(\kk^2) \stackrel{!}=  g_{\alpha}(\kk^2) \stackrel{!}= 0~,
	\ee
	justifying the assumption $g_\beta(\kk^2)\big|_{\vk=0} = g_\beta(\kk^2)\big|_{\vk\rightarrow\infty}= 0$ in \eqref{eq:FS-betalim} and \eqref{eq:FS-betaInf}.

\section{Inverse transform and geodesics}\label{sec:IT}
	Having found the new canonical coordinates $\{\rt,\phit,\pit_r,\pit_\phi\}$ as functions of the old ones $\{r,\phi,\pi_r,\pi_\phi\}$, in this section we invert the map. As at least locally the new coordinates $\rt$ and $\phit$ describe a two-sphere, we then obtain geodesics on the FS by mapping the $\Sph^2$ geodesics back to the FS. Finally, we comment on the closure of orbits and on the range of the angle variable $\alpha$.

\subsection{Inverse transform}
	Recall that $K_A$ are just the isometries of the sphere defined by the new variables $\rt$ and $\phit$\footnote{More precisely, of what would be an $\Sph^2$ if not the range of $\zt=\sqrt{1-\rt^2}$ was too restricted.} and that, by \eqref{eq:FS-calKsq} and \eqref{eq:NCC-KsqAlphaPB}, $\kk$ and $\alpha$ are well understood and well behaved quantities. Since by definition $\pit_\phi = -K_3 =\pi_\phi$ \eqref{eq:NCC-pitphi} and since \eqref{eq:NCC-rtSol2} and \eqref{eq:NCC-pitrSol} establish the map between $\{\rt,\pit_r\}$ and $\{\kk,\alpha\}$, we will derive expressions for the old phase space variables $\{r,\phi,\pi_r,\pi_\phi\}$ in terms of $\{\alpha,\phit,k,\pi_\phi\}$.
	
	Inverting \eqref{eq:NCC-alphaSol} for $\PP$ we get 
	\be
		\label{eq:IT-Pinv}
		\PP(\kk,\alpha)=\sqrt{\frac{\kk^2}{1+\vk^2}}\, \text{sn}\left(\uu| \mm\right)~,
	\ee
	$\mm = \frac{\vk^2\kk^2}{(1+\vk^2)(1+\kk^2)}$ given in \eqref{eq:FS-EllArgs} and $\text{sn}(\uu| \mm)$ etc. being Jacobi elliptic functions.

	Plugging this into \eqref{FS-PPKKinv} we find
	\begin{align}
		\label{eq:IT-rinv}
		&r(\kk,\alpha) = \frac{\sqrt{1+\vk^2 + \kk^2\, \text{sn}\left(\uu| \mm\right)^2}}{\sqrt{1+\vk^2}\sqrt{1+\kk^2}\,\text{dn}\left(\uu| \mm\right)}~,\\
		\label{eq:IT-pirinv}
		&\pi_r(\pi_\phi,\kk,\alpha) = \pi_\phi \left(1+\kk^2\right)\frac{\text{dn}\left(\uu| \mm\right)^2\, \text{sc}\left(\uu| \mm\right)}{\sqrt{1+\vk^2 + \kk^2\, \text{sn}\left(\uu| \mm\right)^2}}~.
	\end{align}

	Furthermore, inserting \eqref{eq:IT-Pinv} into the solution for $\beta(\kk,\PP)$ \eqref{eq:FS-betaSol} we have
	\be
		\label{eq:IT-betainv2}
		\beta(\kk,\alpha) = \frac{1+\vk^2+\kk^2}{\sqrt{1+\vk^2}\sqrt{1+\kk^2}} \,\ellPi\left(\cc ; \text{am}\left(\uu| \mm\right)\big| \mm \right) - \frac{\vk^2}{\sqrt{1+\kk^2}}\,\alpha + g_\beta(\kk)~,
	\ee
	where in \eqref{eq:NCC-galphabeta2} we set $g_\beta(\kk)=0$. By $\phit=\phi -\beta + \arctan(\rt \varpi_r)$ \eqref{eq:NCC-phit} and $\rt \varpi_r = - \sqrt{1+\kk^2} \cot(\alpha)$ \eqref{eq:NCC-rtsq-pitrrt} we therefore have
	\be\ba
		\label{eq:IT-phiinv}
		\phi(\phit,\kk,\alpha) =\,& \phit + \frac{1+\vk^2+\kk^2}{\sqrt{1+\vk^2}\sqrt{1+\kk^2}} \,\ellPi\left(\cc ; \text{am}\left(\uu| \mm\right)\big| \mm \right)\\
		&- \frac{\vk^2}{\sqrt{1+\kk^2}}\,\alpha + \arctan\left(\sqrt{1+\kk^2}\cot(\alpha) \right)\,.
	\ea\ee

\subsection{Generating geodesics on the Fateev sausage}
	With these expressions at hand, we now can generate geodesics on the FS as follows: In terms of new phase space variables, the Hamiltonian \eqref{eq:FS-FSasS2} describes the particle on $\Sph^2$ and geodesics are given by great arcs, {\it i.e.}, maximal circles. Applying the inverse transform above, we then map these great arcs to geodesics on the FS. 

	Therefore, let us start by generating the most general great arcs on $\Sph^2$.
	In analogy to \eqref{eq:FS-S2Embedd}, $\Reals^3$ embedding for half an $\Sph^2$ are given by
	\be
		\label{eq:IT-S2cropEmbedd}
		\vec{\widetilde{X}} = \left(\widetilde{X}^1,\widetilde{X}^2,\widetilde{X}^3\right) = \left(\rt\cos(\phit), \rt\sin(\phit), \sqrt{1-\rt^2}\right)\qquad\text{with}\quad 0 \leq \rt \leq 1~.
	\ee
	As any great arc passes through the equator, we choose this to happen at world-line time $\tau=0$,
	\be
		\label{eq:IT-rttau0}
		\rt\big|_{\tau=0} = 1~,\qquad\qquad \widetilde{X}^3\big|_{\tau=0}=0~.
	\ee
	Next, we use the rotational invariance along the $\phit$-direction to set
	\be
		\label{eq:IT-phittau0}
		\phit\big|_{\tau=0} = 0 ,\qquad\qquad \vec{\widetilde{X}}\big|_{\tau=0} = \big(1,0,0\big)~.
	\ee
	Hence, there are only two parameters left distinguishing different great arcs: Its angle $\th$ relative to the equator and its momentum $\pi_\alpha$ along the great arc. A geodesic on the equator having momentum $\pi_\alpha = \pi_\phi$ and fulfilling $\vec{\widetilde{X}}\big|_{\tau=0} = \big(1,0,0\big)$ is given by
	\be
		\vec{\widetilde{X}} = \big(\!\cos(\pi_\alpha \tau),\sin(\pi_\alpha \tau),0\big)~.
	\ee
	A non-vanishing angle $\th$ relative to the equator is attained by rotation along the $\widetilde{X}^1$-axis, giving
	\be
		\vec{\widetilde{X}} 
		= \big(\!\cos(\pi_\alpha\,\tau ),\cos(\th) \sin(\pi_\alpha\, \tau ),\sin(\th) \sin(\pi_\alpha\, \tau )\big)~,
	\ee
	where $\th \in [-\pi/2,\pi/2]$ to omit double counting of geodesics.
	
	We can calculate the corresponding angular momenta $K_A$ according to \eqref{eq:NCC-S2IsomK}, and find
	\be
		\vec{K}=\big(K_1,K_2,K_3) = \big(0,\pi_\alpha \sin(\th), -\pi_\alpha \cos(\th)\big)~,
	\ee
	hence
	\be
		\th = \arctan(-K_2 /K_3) = \arctan(\kk)~,\qquad\quad \pi_\alpha = \pi_\phi \sqrt{1+\kk^2} = \text{sgn}(\pi_\phi) \abs{\vec{K}} ~,
	\ee
	where we used $K_3 = - \pi_\phi$ and $\th \in [-\pi/2,\pi/2]$. 
	Therefore, apart from the conditions \eqref{eq:IT-rttau0} and \eqref{eq:IT-phittau0}, geodesics on $\Sph^2$ are completely determined by $\pi_\alpha= \pi_\phi \sqrt{1+\kk^2}$ and $\kk$,
	\be
		\label{eq:IT-XtGeod}
		\vec{\widetilde{X}} 
		= \left(\cos\big(\pi_\alpha\,\tau\big),\frac{1}{\sqrt{1+\kk^2}}\,\sin\big(\pi_\alpha\,\tau\big) ,\frac{\kk}{\sqrt{1+\kk^2}}\, \sin\big(\pi_\alpha\,\tau\big) \right)~.
	\ee
	
	Calculating $\rt = \sqrt{1-\big(\widetilde{X}^3\big){\vphantom{X}}^2}$ and its conjugate momentum $\pit_r = \frac{1}{1-\rt^2} \p_\tau \rt$ we have
	\begin{align}
		&\rt = \sqrt{1 - \frac{\kk^2 \sin^2\big(\pi_\alpha\,\tau\big)}{1+\kk^2}}
			= \sqrt{\frac{1 + \kk^2 \cos^2\big(\pi_\alpha\,\tau\big)}{1+\kk^2}}~,\\
		&\pit_r = -\pi_\alpha \sqrt{\frac{1+\kk^2}{1+\kk^2 \cos^2(\pi_\alpha\,\tau)}} \cot(\pi_\alpha\,\tau)
			= \frac{-\pi_\phi (1+k^2)\cot(\pi_\alpha\,\tau)}{\sqrt{1+\kk^2 \cos^2(\pi_\alpha\,\tau)}} ~,
	\end{align}
	and by comparison with \eqref{eq:NCC-rtSol2} and \eqref{eq:NCC-pitrSol} we see that
	\be
		\alpha = \pi_\alpha\,\tau~.
	\ee
	
	Indeed, this fits together with the observation in \eqref{eq:NCC-KsqAlphaPB} that $\pi_\alpha = \pi_\phi \sqrt{1+\kk^2}$ is the momentum canonically conjugate to $\alpha$, which justifies our notation and at the same time proves that \eqref{eq:IT-XtGeod} is a geodesic on $\Sph^2$. Furthermore, we see that $\kk$ and $\alpha = \pi_\alpha \tau$, together with $\pi_\phi$ and $\phit$, are much more convenient variables to describe geodesics than $\rt$ and $\pit_r$, justifying the form of the inverse transformations \eqref{eq:IT-Pinv} to \eqref{eq:IT-phiinv}.

	Finally, on the great arc we find for $\phit$
	\be
		\phit = \text{arccot}\left(\widetilde{X}^1/\widetilde{X}^2\right) = \text{arccot}\left(\sqrt{1+\kk^2}\,\text{cot}(\pi_\alpha\,\tau)\right) = \frac{\pi}{2}+\arctan( \rt \varpi_r)~,
	\ee
	where we used \eqref{eq:NCC-rtsq-pitrrt}. Therefore, on the great arc \eqref{eq:IT-phiinv} simplifies to
	\be
		\label{eq:IT-phiinv2}
		\phi(\kk,\alpha) = \frac{1+\vk^2+\kk^2}{\sqrt{1+\vk^2}\sqrt{1+\kk^2}} \,\ellPi\left(\cc ; \text{am}\left(\uu| \mm\right)\big| \mm \right)
		- \frac{\vk^2}{\sqrt{1+\kk^2}}\,\alpha + \frac{\pi}{2}~.
	\ee

	\begin{figure}[ht]
		\centering
		\qquad\begin{minipage}[b]{0.45\linewidth}
		\includegraphics[width=7cm]{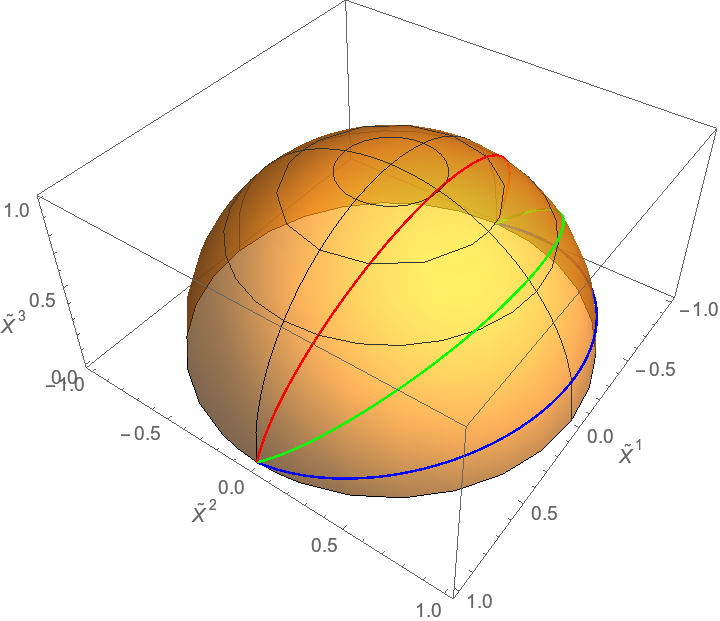}
		\end{minipage}
		\quad 
		\begin{minipage}[b]{0.45\linewidth}
		\includegraphics[width=7cm]{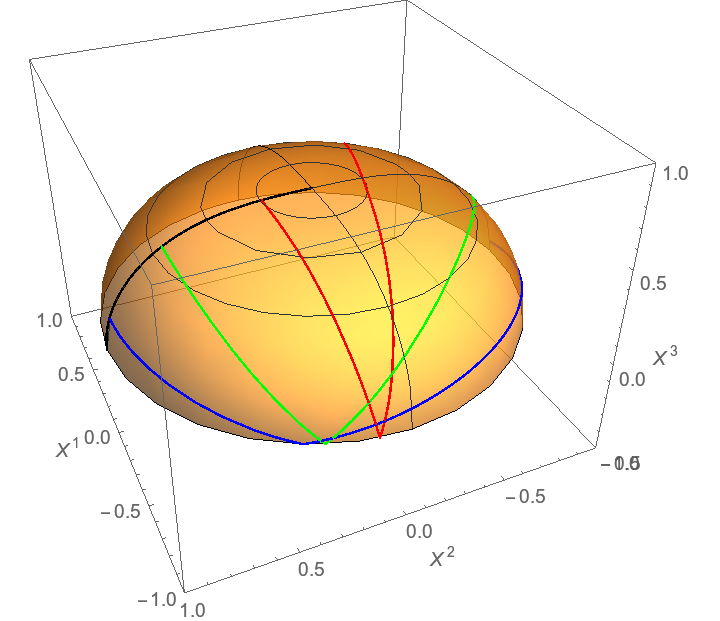}
		\end{minipage}
		\vspace{.5cm}\\
		\qquad\begin{minipage}[b]{0.45\linewidth}
		\includegraphics[width=7cm]{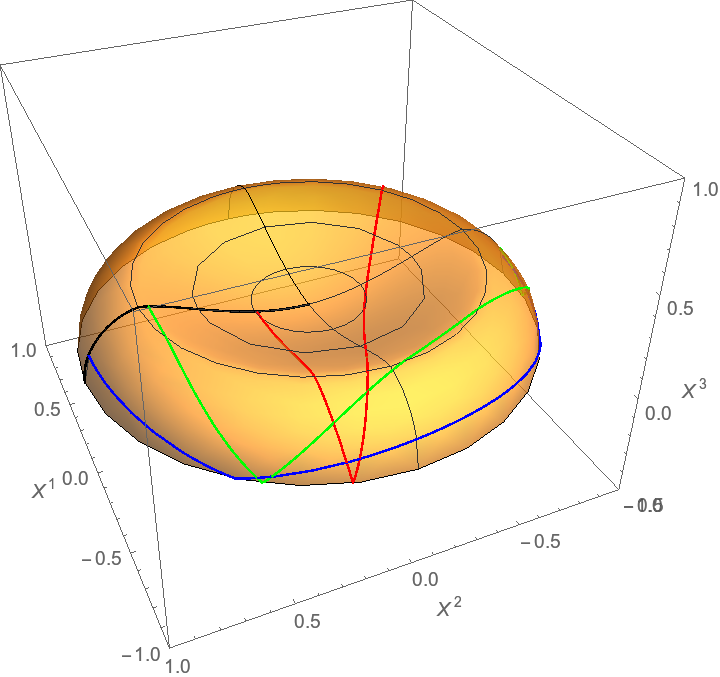}
		\end{minipage}
		\quad 
		\begin{minipage}[b]{0.45\linewidth}
		\includegraphics[width=7cm]{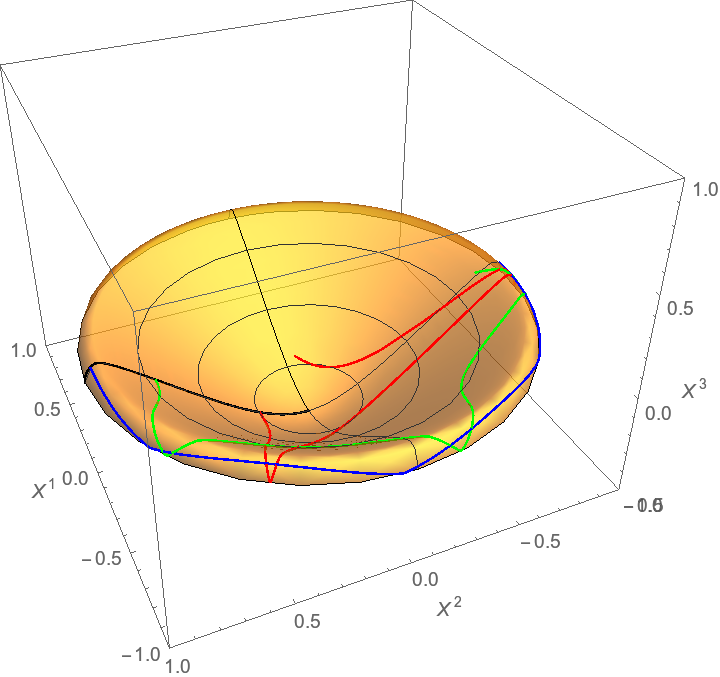}
		\end{minipage}
		\quad
		\caption{Geodesics for $\alpha\in[0,\pi]$ and $\kk=1/4$ (blue), $\kk=1$ (green), and $\kk=4$ (red) on the $\Sph^2$ defined by $\rt$ and $\phit$ (u.l.), and on the FS defined by $r$ and $\phi$ for $\vk=1$ (u.r.), $\vk=1.5$ (l.l.), and $\vk=3$ (l.r.).}
		\label{fig:geodesics}
	\end{figure}

	We have now everything to generate geodesics on the FS and corresponding plots (for embedding coordinates \eqref{eq:FS-FSembedd3}) are shown in Figure \ref{fig:geodesics}. There, great arcs on half the $\Sph^2$ defined by the new coordinates $\rt$ and $\phit$ (upper left) for $\alpha = \pi_\alpha \tau \in [0,\pi]$ and $\kk=\frac{1}{4}$ (blue), $\kk=1$ (green), and $\kk=4$ (red) are mapped to half the FS defined by the original coordinates $r$ and $\phi$ for the deformation parameter taking values $\vk=1$ (upper right), $\vk=1.5$ (lower left), and $\vk=3$ (lower right).

	We observe that the starting point $\vec{\widetilde{X}}\big|_{\tau=0} = \big(1,0,0\big)$ on the $\Sph^2$ is mapped to different points on the meridian formed by the intersection of the FS with the $X^2$-$X^3$-plane, depicted as thick black line. This is a manifestation of the coordinates $r$ and $\phi$ containing information on the momenta, in particular on $\kk$.

\subsection{Closure of orbits and range of the transformation}\label{subsec:NCC-Range}
	In Figure \ref{fig:geodesics}, we notice moreover that the generated geodesics on half the FS are reflected at the boundary $r=1$. But of course, this is an artifact of our parametrization covering only half the FS and we can use this feature to recover geodesics on the whole FS.

	For this, recall that $r(k,\alpha)$ \eqref{eq:IT-rinv} depends on the angle variable $\alpha$ along the geodesic via Jacobi elliptic functions, in particular $\text{sn}(\uu, \mm)$ and $\text{dn}(\uu, \mm)$. 
	As the modulus squared $\mm$ \eqref{eq:FS-EllArgs} is bound by zero and one,
	\be
		\label{eq:IT-mBound}
		0\leq m =\frac{\vk^2 k^2}{(1+\vk^2)(1+\kk^2)} \leq 1~,
	\ee
	these Jacobi elliptic functions are periodic and undergo a full cycle for a shift in the angle variable $\alpha$ along the geodesic by
	\be
		\label{eq:IT-DeltaAlpha}
		\Delta\alpha(\kk) = \frac{4}{\sqrt{1+\vk^2}}\,\ellK(\mm) = \frac{4}{\sqrt{1+\vk^2}}\,\ellK\left(\frac{\vk^2 \kk^2}{(1+\vk^2)(1+\kk^2)}\right)~.
	\ee

	As the geodesics start on a meridian, see Figure \ref{fig:geodesics}, by reflection symmetry we conclude that the geodesics hit the boundary at the values
	\be
		\label{eq:IT-alphaS}
		\alpha_s(\kk) = \frac{2s +1}{4} \Delta{\alpha}(\kk) = \frac{2s + 1}{\sqrt{1+\vk^2}}\,\ellK\left(\frac{\vk^2 \kk^2}{(1+\vk^2)(1+\kk^2)}\right)~,\qquad s\in\Integers~.
	\ee
	We therefore obtain geodesics on the full FS by alternating at these $\alpha_s(\kk)$ between an upper and a lower half FS, {\it i.e.}, effectively by alternating the sign of the $X^3$ component of the embedding coordinates \eqref{eq:FS-FSembedd3}.

	\begin{figure}[ht]
		\centering
		\begin{minipage}[b]{0.45\linewidth}
		\includegraphics[width=7cm]{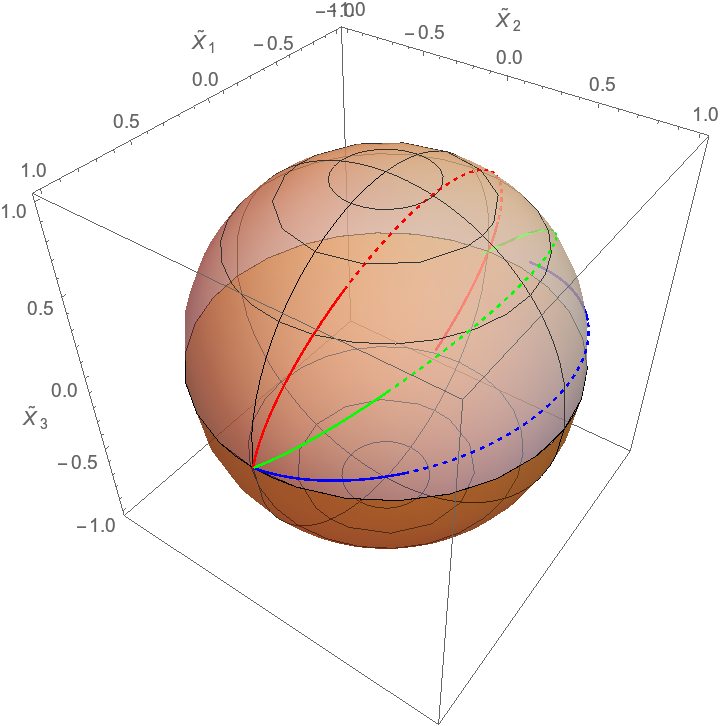}
		\end{minipage}
		\qquad 
		\begin{minipage}[b]{0.45\linewidth}
		\includegraphics[width=7cm]{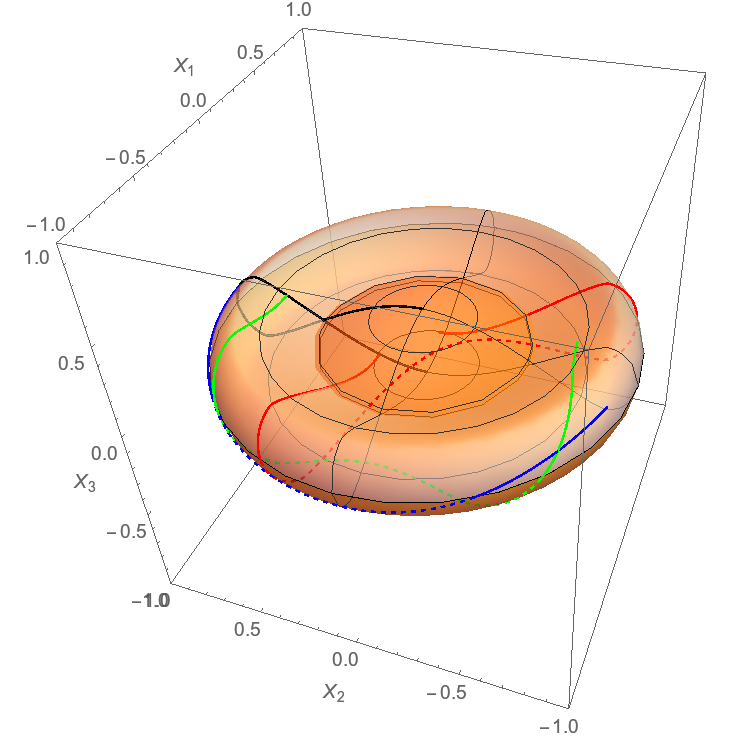}
		\end{minipage}
		\caption{Full geodesics for $\kk=1/4$ (blue), $\kk=1$ (green), and $\kk=4$ (red) for $\alpha\in[0,\Delta\alpha(\kk)]$ on the $\Sph^2$ defined by $\rt$ and $\phit$ (left), and on the FS defined by $r$ and $\phi$ for $\vk=2$ (right).}
		\label{fig:geodesicsFull}
	\end{figure}

	An example is shown in Figure \ref{fig:geodesicsFull}. There, geodesics with $\kk=1/4$ (blue), $\kk=1$ (green), and $\kk=4$ (red) are mapped from the sphere of new coordinates (left) to the FS for $\vk=2$ (right). Each geodesic is traced out for $\alpha\in[0,\Delta\alpha(\kk)]$, where $\alpha\in\big[\alpha_1(k),\alpha_2(k)\big]=\big[\frac{\Delta\alpha(\kk)}{4},\frac{3 \Delta\alpha(\kk)}{4}\big]$ are depicted dashed, as these are the intervals for which the geodesics lie on the lower FS.

	We observe that at the end points the geodesics are again parallel to the equator, showing that $r(\alpha,\kk)$ underwent a full cycle. Note however, that the end points lie at different angles $\phi(\alpha,\kk)$.

	Correspondingly, although the great arcs on the $\Sph^2$ defined by $\rt$ and $\phit$ close, generally we can not expect geodesics to close on the FS. We can quantify this suspicion by calculating the difference in the angle $\phi(\alpha,\kk)$ after one cycle of $r(\alpha,\kk)$. Plugging \eqref{eq:IT-DeltaAlpha} into \eqref{eq:IT-phiinv2} we find
	\be
		\Delta\phi(k) = \phi(\Delta\alpha,k) - \phi(0,k) = \frac{4(1+\vk^2+k^2)}{\sqrt{1+\vk^2}\sqrt{1+k^2}}\,\ellPi\big(\cc;\mm\big) 
		-\frac{4 \vk^2}{\sqrt{1+\vk^2}\sqrt{1+k^2}}\,\ellK\big(\mm\big)~.
	\ee
	Hence, for fixed deformation parameter $\vk$ an orbit of inclination $\th = \arctan(\kk)$ closes for $\Delta\phi(\kk)$ being a rational multiple of $2\pi$. As (for $\vk\neq0$) $\Delta\phi(\kk)$ appears to be a smooth, non-constant function of $\kk\in\Reals_+$, this is the case for almost no geodesic in the sense of $\mathbb{Q}$ having Lebesgue measure zero.
	Put differently, the motion on the FS is generally not periodic but only quasi-periodic as the frequency $1/\Delta\alpha(k)$ is incommensurable with $1/2\pi$, the frequency of $\phi$. In accordance, the canonical transformation of Section \ref{sec:NCC} does not map geodesics on the FS $(\Sph^2)_\eta$ to a two-sphere $\Sph^2$, but rather to its covering space, with the angle variable $\alpha$ decompactified. 

\begin{figure}[ht]
		\centering
		\begin{minipage}[b]{0.5\linewidth}
		\includegraphics[width=8cm]{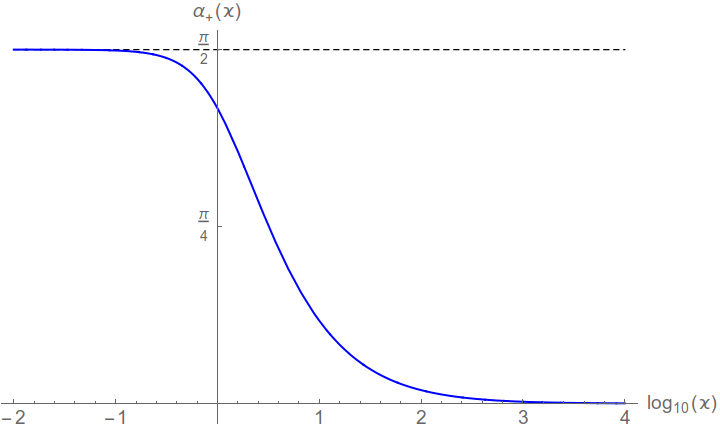}
		\end{minipage}
		\caption{Maximal value $\alpha_+(\vk)$ of $\alpha(\kk,\PP)$ as a function of $\log_{10}(\vk)$.}
		\label{fig:alphaMax}
	\end{figure}

	Finally, with the above discussion at hand we can attack an open end in Section \ref{sec:NCC}. At several occasions in the derivation of the new canonical coordinates we assumed $\alpha \in [0,\pi/2]$ \eqref{eq:NCC-alphaInterval}, while when applying the inverse transform to obtain geodesics on the FS we did not run into any problems for arbitrary $\alpha$.
	To understand this mismatch, we calculate the range of $\alpha(\kk,\PP)$ \eqref{eq:NCC-alphaSol} for all possible values of old phase space variables, that is, for all $\kk$ and $\PP$. As for $0\leq m \leq 1$ \eqref{eq:IT-mBound} the elliptic integral $\ellF(\Phi|m)$ in \eqref{eq:NCC-alphaSol} is monotonically increasing, $\alpha(\kk,\PP)$ takes its maximal and minimal values for $\PP$ satisfying the inequality \eqref{eq:FS-ineq}, $p \rightarrow \pm\sqrt{k^2/(1+\vk^2)}$, giving $\alpha(\kk,\pm\infty) = \pm\frac{\Delta\alpha}{4} = \pm \alpha_1(\kk)$. Hence, by \eqref{eq:IT-alphaS} this corresponds to points on the boundary of half a FS, $r=1$. The extremal values of $\alpha(k,\pm\infty)$ are then reached by taking $\kk\rightarrow\infty$, yielding
	\be
		\label{eq:IT-alphapm}
		\alpha_{\pm}(\vk) = \pm \alpha_1(\kk)\Big|_{\kk\rightarrow\infty} = \frac{\pm 1}{\sqrt{1+\vk^2}}\,\ellK\left(\frac{\vk^2}{1+\vk^2}\right)~,
	\ee
	with the corresponding plot of $\alpha_+(\vk)$ over $\log_{10}(\vk)$ given in Figure \ref{fig:alphaMax}.

	We read off that $\alpha(\kk,\PP)$ occupies the interval $[\alpha_-(\vk),\alpha_+(\vk)] \subset [-\pi/2,\pi/2]$. Especially, the interval $[\alpha_-(\vk),\alpha_+(\vk)]$ is almost $[-\pi/2,\pi/2]$ for small $\vk$, while for $\vk$ large it diminishes quickly. To bring this into agreement with \eqref{eq:NCC-alphaInterval}, we note that we could have allowed for $\alpha \in [-\pi/2,\pi/2]$ by instead of \eqref{eq:IT-pitrSol} taking $\pit_r(\kk,\alpha) \mapsto \text{sign}(\alpha)\pit_r(\kk,\alpha)$~,
	\be
		\label{eq:IT-pitrSol}
		\pit_r(\kk,\alpha)= -\pi_\phi \frac{(1+\kk^2) \sqrt{\cot^2(\alpha)}}{\sqrt{1+\kk^2 \cos^2\left(\alpha\right)}}~.
	\ee
	
	The interpretation of the above behaviour is simply that $\alpha(\kk,\PP)$ \eqref{eq:NCC-alphaSol} is multi-valued as a function of old phase space variables, whereas the inverse transform, in particular $\PP(\kk,\alpha)$ \eqref{eq:IT-Pinv}, is single valued, which allowed us to generate geodesics on the FS easily for arbitrary $\alpha$. Of course, these are just the properties of the elliptic integrals and their inverse, the Jacobi elliptic functions.

\section{Outlook}\label{sec:Outlook}
	In this paper we have demonstrated the maximal superintegrability of geodesic motion in the FS model, {\it i.e.}, on $(\Sph^2)_\eta$, which is also a particular reduction of the integrable sigma model on $(\ads)_{\eta}$. 
	The corresponding dynamics is governed by the Hamiltonian of the two-dimensional $\eta$-deformed Rosochatius system for which we 
	exhibited three non-abelian integrals of motion $K_A$, yielding corresponding integrals of motion $J_{\alg{m},A}$ for the mirror geometry of $\rm{S}^2$ \cite{Arutyunov:2014cra} as a by-product. We have also found a canonical transformation which maps geodesics on the auxiliary two-sphere to the ones on the FS manifold. In spirit, our method seems to be related to the mapping of geodesics to an auxiliary sphere developed for the spheroid by the likes of Legendre, Oriani, and Bessel \cite{Legendre:1806,Oriani,Bessel,Bessel2}. The connection is though disguised by the use of different languages, Hamiltonian mechanics in our case versus differential geometry for \cite{Legendre:1806,Oriani,Bessel,Bessel2}.

	We pointed out that our method applies to arbitrary two-dimensional surfaces of revolution, as is the case for \cite{Legendre:1806,Oriani,Bessel,Bessel2}, and it is intriguing to investigate other cases of interest. In particular, it would be worthwhile to see whether our formulas still apply for the case 
	of the complex deformation parameter $\vk$, see \eqref{eq:FS-vknu}, where the FS manifold indeed looks like a sausage rather than a squashed sphere as we have in our present case.

	Another pressing question is the geodesic problem on $(\AdS_2)_\eta$, which exhibits a naked singularity. Special cases for geodesics have been studied previously in \cite{Kameyama:2014via} and in the unpublished work by S. Frolov and R. Roiban. Due to the similarity of $(\AdS_2)_\eta$ and $(\Sph^2)_\eta$ it is tempting to hope that the general solution for geodesics on $(\AdS_2)_\eta$ can be obtained by minor modifications of the presented formulae. The corresponding integrals for the mirror geometry of $\AdS_2$ \cite{Arutyunov:2014cra} could then again be found by taking appropriate limits.

	Certainly, the program pursued in this work has the final goal of understanding point particle solutions in the $\eta$-deformed $\AdS_5 \times \Sph^5$ sigma model. Since in the absence of fermions the sigma model on $(\ads)_{\eta}$ factorizes into the product of the $({\rm AdS}_5)_{\eta}$ and $({\rm S}^5)_{\eta}$ models it would be interesting to investigate the question whether the five-dimensional $\eta$-deformed Rosochatius model, which describes geodesics on $({\rm S}^5)_{\eta}$, also has a chance to exhibit maximal superintegrability. We expect however, that the answer for this bosonic five-dimensional model is obscured by the fact, that the additional integral of motion $Q$ found in \cite{Arutyunov:2014cda, Arutyunov:2016ysi} is not an isometry anymore. It might be therefore advisable to first disentangle the geodesic problem for the submanifold obtained for zero momenta along the isometries $\phi_{1,2,3}$. 

	Another stepping stone would be to investigate geodesic motion on $\eta$-deformed $\Sph^3$ and $\AdS_3$, with the former corresponding to the so-called Fateev ${\rm O}(4)$ model, see \cite{Fateev:1996ea} and \cite{Lukyanov:2012zt}. As in comparison to the present case of geodesics on $(\Sph^2)_\eta$ only one more isometry is added, we are optimistic that the derived method can be generalized to this setting.  
	
	Moreover, one has to wonder if the found superintegrability and canonical transformation for geodesic motion have interesting implications for other string-like solutions and for $\eta$-deformed sigma models in general. 
	In particular, in \cite{Kyono:2015iqs} the $\eta$-deformation\footnote{Here, the term ``$\eta$-deformed'' stems from the fact that the construction closely follows the one for the $(\AdS_5 \times \Sph^5)_\eta$ sigma model, although its notion is questionable due to the lack of integrability.} of the Nappi-Witten model was undone by finding an appropriate coordinates transformation while in \cite{Borsato:2016ose} unimodular deformations were removable by a chain of non-commutative TsT-transformations and non-linear field redefinition.\footnote{See the talk by A. Tseytlin at IGST 2016, Berlin.}
	Although our canonical map from geodesics on $(\Sph^2)_\eta$ to geodesics on $\Sph^2$ could not be constructed globally, as is eminent from the non-closure of geodesics on the FS manifold, these works spark the hope that such symplectic methods find application also for the $(\AdS_5 \times \Sph^5)_\eta$ sigma model.

	Finally, there are two more pressing questions left unanswered. On the one hand, the found integrals $K_A$ only showed to form an $\so(3)$ algebra while we were not able to argue for a quantum deformed isometry algebra corresponding to the quantum group $\,\mathcal{U}_q\big(\so(3)\big)$, as expected by \cite{Delduc:2014kha}. On the other hand, the non-closure of the geodesics stems a fundamental obstacle to quantization of the system. In particular, the particle on $(\Sph^2)_\eta$ is quantum mechanically different from the one on $\Sph^2$ and it would be worthwhile to determine its spectrum.
	Insights into both these problems might be devised by use of the Kirillov-Kostant-Souriau method of coadjoint orbits, see also the recent works \cite{Heinze:2015oha, Heinze:2016fin, Heinze:2016lxs}, as its extension to quantum groups has been investigated \cite{KirillovMerits}. 

	A different explanation to the first problem might be that geodesic motion does not entail rich enough world-sheet dynamics and that to observe the Hopf-algebra structure of a quantum deformed isometry algebra one has to study more involved string-like solutions. For instance, it would be interesting to see whether also for the $\eta$-deformed Neumann \cite{Arutyunov:2014cda} and Neumann-Rosochatius models \cite{Arutyunov:2016ysi} additional non-abelian integrals of motion can be constructed.

\subsection*{Acknowledgements}
\noindent
We are grateful to Riccardo Borsato, Amit Dekel, Ben Hoare, George Jorjadze, Radu Roiban, Stijn van Tongeren, Kentaroh Yoshida, Konstantin Zarembo and Alexander Zheltukhin for useful discussions.
We also thank Arkady Tseytlin and Stijn van Tongeren for comments on the manuscript. 
M.H. thanks NORDITA for hospitality.
The work of G.A. and M.H. is supported by the German Science Foundation (DFG) under the Collaborative Research Center (SFB) 676 Particles, Strings and the Early Universe. The work of D.M. is supported by the ERC advanced grant No 341222.

\appendix
\section{Integrals of motion for the spheroid}\label{app:Spheroid}
	In the introduction and the conclusion we have asserted that the method for construction of new integrals of motion developed in the present work is generally applicable to geodesic motion on any two-dimensional surface of revolution. To illustrate this point, in this appendix we derive the corresponding integrals of motion $K_A$ for the spheroid\footnote{We do not invent separate notation for spheroid quantities, as many expressions stay unaltered.}, {\it i.e.}, for the ellipsoid of revolution.

  For definiteness, let us consider the spheroid with polar semi-axes of length $c$ and equatorial radius set to one. Corresponding $\Reals^3$ embedding coordinates are $\vec{X}=(r \cos(\phi),r\sin(\phi),c \sqrt{1-r^2})$, yielding for geodesic motion the Hamiltonian
	\be
		\label{eq:App-Ham}
		H = \frac{1-r^2}{2\big((1-r^2)+c^2 r^2\big)}\pi_r^2 + \frac{\pi_\phi^2}{2 r^2}~.
	\ee
	
	The momentum $\pi_\phi$ is an integral of motion, as required, and as in \eqref{eq:FS-K3} we set $K_3 = -\pi_\phi$. As the Hamiltonian \eqref{eq:App-Ham} is quadratic in momenta, still the same arguments apply and for $K_{A=1,2}$ we again take the ansatz \eqref{eq:FS-K12-3}.
	
	The only difference to Section \ref{sec:PIFS} lies now in the choice of appropriate variables $\kk$ and $\PP$, see \eqref{eq:FS-PPkkdef}, and the resulting ODE for $\beta(\kk,\PP)$ \eqref{eq:FS-betaODE}. In particular, defining
	\be
		\label{eq:App-PPKKdef}
		\PP = \frac{r}{\sqrt{(1-r^2)+c^2 r^2}}\,\frac{\pi_r}{\pi_\phi}~,\qquad
		\kk = \sqrt{\frac{2 H - \pi_\phi^2}{\pi_\phi^2}}
			= \sqrt{1-r^2} \sqrt{\frac{1}{r^2}+ \frac{1}{(1-r^2)+c^2 r^2}\,\frac{\pi_r^2}{\pi_\phi^2}}~,
	\ee
	and imposing $\{H,K_A\}_{\rm PB}=0$ we again find \eqref{eq:FS-ffcsODEs}, where now however
	\be
		\label{eq:App-Cbeta}
		C_\beta(\kk,\PP) = \frac{(1+\PP^2)\sqrt{1+\kk^2 + \PP^2}}{\sqrt{\kk^2 +c^2 (1+\PP^2)}}~.
	\ee
	
	Solving \eqref{eq:FS-betaODE}, $\p_\PP \beta(\kk,\PP) = C_\beta(\kk,\PP)^{-1}$, we find the solution
	\be
		\label{eq:App-betaSol}
		\beta(\kk,\PP) = \frac{c^2-1}{\sqrt{c^2 + \kk^2}}\,\ellF\left(\Phi,\mm\right) 
			+ \frac{1+\kk^2}{\sqrt{c^2 + \kk^2}}\,\ellPi\left(-\kk^2,\Phi,\mm \right)+g_\beta(k)~,
	\ee
	with the argument $\Phi$ and the modulus squared $\mm$ now given by
	\be
		\Phi = \arctan\left(\frac{\PP}{\sqrt{1+\kk^2}}\right)~,\qquad\quad \mm=\frac{(1-c^2)\kk^2}{c^2 + \kk^2}~.
	\ee

	Hence, the final form of the charges $K_{A=1,2}$ is given again by \eqref{eq:FS-K12final} (but with $\PP$ and $\kk$ in \eqref{eq:App-PPKKdef} and $\beta(\kk,\PP)$ in \eqref{eq:App-betaSol}), where the choice \eqref{eq:FS-gAcs} results in an $\so(3)$ algebra \eqref{eq:FS-PBKK}.

	We conclude by noting that the above expressions simplify dramatically for $c=1$. Of course, this has to be expected as it corresponds to the particular case of the spheroid being an $\Sph^2$.

\bibliographystyle{nb}
\bibliography{bibSpires}

\end{document}